\documentclass[twocolumn]{autart}    

\makeatletter
\def\ps@pprintTitle{%
	\let\@oddhead\@empty
	\let\@evenhead\@empty
	\def\@oddfoot{\centerline{\thepage}}%
	\let\@evenfoot\@oddfoot}
\makeatother

\usepackage{graphicx}          
                               
\usepackage{cite}
\usepackage[noprefix]{nomencl}
\usepackage{amsmath} 
\usepackage{amssymb}  

\makenomenclature
\begin{document}

\begin{frontmatter}

\title{Model Reference Adaptive Control Allocation for Constrained Systems with Guaranteed Closed Loop Stability}


\author[Turkey]{Seyed Shahabaldin Tohidi}\ead{shahabaldin@bilkent.edu.tr},    
\author[Turkey]{Yildiray Yildiz}\ead{yyildiz@bilkent.edu.tr},               
\author[USA]{Ilya Kolmanovsky}\ead{ilya@umich.edu}  

\address[Turkey]{Mechanical Engineering Department, Bilkent University, Ankara, 06800, Turkey}             
\address[USA]{Department of Aerospace Engineering, University of Michigan, Ann Arbor, MI 48109, USA}        


\begin{abstract}                          
This paper proposes an adaptive control allocation approach for uncertain over-actuated systems with actuator saturation. The proposed method does not require uncertainty estimation or a persistent excitation assumption. Using the element-wise non-symmetric projection algorithm, the adaptive parameters are restricted to satisfy certain optimality conditions leading to overall closed loop system stability. Furthermore, a sliding mode controller with a time-varying sliding surface, working in tandem with the adaptive control allocation, is proposed to guarantee the outer loop stability and reference tracking in the presence of control allocation errors and disturbances. Simulation results are provided, where the Aerodata Model in Research Environment is used as an over-actuated system with actuator saturation, to demonstrate the effectiveness of the proposed method.\end{abstract}

\end{frontmatter}

\nomenclature[1.1]{$\theta_v$}{Adaptive parameter matrix.}
\nomenclature[1.2]{$\theta_{v_{j}}$}{The $ j $th column of $\theta_v$.}
\nomenclature[1.3]{$\theta_{v_{i,j}}$}{The element at the $ i $th row and $ j $th column of $\theta_v$.}
\nomenclature[1.4]{$\theta_{min_{i,j}}$}{The minimum value of $\theta_{v_{i,j}}$, assigned by the projection algorithm.}
\nomenclature[1.5]{$\theta_{max_{i,j}}$}{The maximum value of $\theta_{v_{i,j}}$, assigned by the projection algorithm.}
\nomenclature[2.1]{$\theta_v^*$}{Ideal parameter matrix.}
\nomenclature[2.2]{$\theta_{v_{i,j}}^*$}{The element at the $ i $th row and $ j $th column of $\theta_v^*$.}
\nomenclature[2.3]{$\theta_{min_{i,j}}^*$}{The minimum value of $\theta_{v_{i,j}}^*$.}
\nomenclature[2.4]{$\theta_{max_{i,j}}^*$}{The maximum value of $\theta_{v_{i,j}}^*$.}
\nomenclature[2.5]{$\theta_I^*$}{$\theta_v^*$ when $ \Lambda=I $.}
\nomenclature[2.6]{$\theta_{I_{i,j}}^*$}{The element at the $ i $th row and $ j $th column of $\theta_I^*$.}
\nomenclature[3.1]{$\tilde{\theta}_v$}{The deviation of the adaptive parameter matrix from the ideal parameter matrix.}
\nomenclature[3.2]{$\tilde{\theta}_{v_{i,j}}$}{The element at the $ i $th row and $ j $th column of $\tilde{\theta}_v$.}
\nomenclature[4]{$\tilde{\theta}_{max}$}{The upper bound of the norm of $ \tilde{\theta}_v $ when $ \theta_{v_{i,j}}^* \in [\theta_{min_{i,j}}\ \ \theta_{max_{i,j}}] $.}
\nomenclature[5]{$\tilde{\theta}_{MAX}$}{The upper bound of the norm of $ \tilde{\theta}_v $ when $ \theta_{v_{i,j}}^* \notin [\theta_{min_{i,j}}\ \ \theta_{max_{i,j}}] $.}
\nomenclature[5.1]{$\zeta_{i,j}$}{The projection tolerance of the element at the $ i $th row and $ j $th column of ${\theta}_v$.}
\nomenclature[5.2]{$M_i$}{The bound on the $ i $th virtual control signal $ v_i $.}
\nomenclature[5.3]{$\rho$}{The vector of upper bound of the the elements of the disturbance vector $ d $, denoted as $ \rho=[\rho_1, ..., \rho_r]^T $.}
\nomenclature[5.4]{$\bar{\rho}_i$}{The upper bound of the norm of the $ i $th row of the multiplicative uncertainty ($ \Delta B $).}

	\printnomenclature 

\section{Introduction}
Control allocation is the process of distributing control signals among redundant actuators. Thanks to the benefits of actuator redundancy in systems, such as improved maneuverability, flexibility and fault tolerability, in addition to the decrease in actuator costs due to advances in microprocessors and actuator miniaturization, the number of applications of control allocation has been growing in recent years in such domains as aircraft, spacecraft, unmanned air vehicles \cite{Yil11a, Duc09, Bod02, Lia10, She15, Yil11b, AcoYil14, She17, SadChaZhaThe12, TohYil18}, ships, underwater vehicles \cite{Pod01, Gie06, Joh08, Che13, Sor11, CorCri16}, automobiles \cite{TjoJoh10, Dem13}, robots \cite{Tag11}, and power systems \cite{Bou17, Rao17}.

Control allocation methods can generally be categorized into the following three sets: Pseudo-inverse-based methods, optimization based methods and dynamic control allocation. Given a mapping between a virtual control input $ v $ and the actuator input vector $ u $ defined as $ Bu=v $, in pseudo inverse based control allocation \cite{Dur93, Dur17, Alw08, Toh16}, the control input is distributed to the individual actuators by the pseudo inverse of this mapping $ u=B^+v $. It is known that this distribution minimizes the $ 2 $-norm of the actuator input vector. This approach can be extended to account for actuator saturation \cite{Dur93, Dur17, Toh16, KirSteHor18}. Daisy chaining \cite{Buf96} and redistributed pseudo inverse \cite{Bor96, Vir94, Ste17} are the other modified versions of pseudo inverse method that consider actuator constraints. In optimization based control allocation \cite{PetBod06, Har05, CasGar10, Har02, YilKol10, YilKolAco11}, control input is distributed by minimizing the cost function $ |Bu-v| + J_0 $, where $ J_0 $ refers to secondary objectives such as minimizing actuator deflections. In dynamic control allocation \cite{Zac09, TjoJoh08, FalHol16, TohYil16, TohYil17, Gal18}, the control signals are distributed among actuators using a set of rules dictated by differential equations. 
A survey on control allocation methods can be found in \cite{JohFos13}.

Control allocation is an appealing approach for the design of active fault-tolerant control systems \cite{Duc09, Zha08, Edw10, Sor17}. Optimization based control allocation is used in \cite{TjoJoh10} to improve the performance of steering in faulty automotive vehicles. In another study \cite{Pod01}, thruster forces of an autonomous underwater vehicle are allocated among redundant thrusters using control allocation so that faults are accommodated. In \cite{SadChaZhaThe12}, experimental results are reported demonstrating the redistribution of the control effort, after a fault, among the redundant actuators of a quadrotor helicopter. In several applications, fault detection and isolation methods are employed in parallel with control allocation \cite{Duc09}. In others, faults are assumed to be estimated a priori. In \cite{Alw08}, a sliding mode controller is coupled with a pseudo inverse based control allocation to obtain a fault tolerant controller wherein faults are assumed to be estimated. Similarly in \cite{Sor17}, it is assumed that there exists a fault detection and isolation scheme which is able to estimate and identify stuck-in-place, hard-over, loss of effectiveness and floating actuator faults. In \cite{CriJoh14}, an unknown input observer is applied to identify actuator and effector faults. A fault detection and isolation method, for nonlinear systems with redundant actuators, by using a family of unknown input observers is proposed in \cite{CriJoh15}. In \cite{Toh16} and \cite{CasGar10}, faults are estimated adaptively using a recursive least square method, and an online dither generation method is proposed to guarantee the persistence of excitation.

%

This paper proposes an adaptive control allocation method for uncertain systems with redundant actuators in presence of actuator saturation. The method builds upon successful approaches mentioned above by eliminating the need for uncertainty estimation, and therefore it does not require persistence of excitation. Furthermore, in the proposed approach, a closed loop reference model \cite{GibAnnLav15} is employed for fast convergence without inducing undesired oscillations. The stability of the overall closed loop system, including the controller, control allocator and the plant is rigorously studied. Preliminary results of this study were previously presented in \cite{TohYil16} and \cite{TohYil17}. A modified version of the adaptive control allocation method based on reducing the difference between the derivative of virtual and actual control signals was introduced in \cite{TohYil18}. It was demonstrated in \cite{TohYil18} that the proposed approach can mitigate pilot induced oscillations. In this paper, we provide the complete picture with an overall closed loop stability proof in the presence of actuator saturation, which was missing in these earlier studies.


Other adaptive approaches to control allocation have been described in \cite{TjoJoh08} and \cite{FalHol16}. Different from these approaches, we explicitly consider the actuator saturation, where we guarantee that the control signals remain within their limits all the time, which allows a systematic design of the outer loop controller without assuming its existence a priori.



Apart from the contributions to the control allocation literature mentioned above, we also showed that it is possible to employ the projection algorithm \cite{LavGib11} in a stable manner even if the ideal adaptive parameters are not inside the projection boundaries. To the best of our knowledge, this result was not reported earlier in the literature.

This paper is organized as follows. Section II introduces notations and preliminary results. Section III presents the uncertain over-actuated plant dynamics and the proposed model reference adaptive control allocation approach with a closed loop reference model. A discussion of actuator saturation and its effects on virtual control limits together with the projection algorithm are given in Section IV. The controller design, producing the virtual control input, is presented in section V. The ADMIRE model is used in Section VI to demonstrate the effectiveness of the proposed approach in the simulation environment. Finally, a summary is given in Section VII.

\section{Background}
In this section, we collect several definitions and basic results which are exploited in the following sections. Throughout this paper, $ ||.|| $ refers to the Euclidean norm for vectors and induced 2-norm for matrices, and $ ||.||_F $ refers to the Frobenius norm.

\begin{figure}
	\begin{center}
		\includegraphics[width=8.5cm]{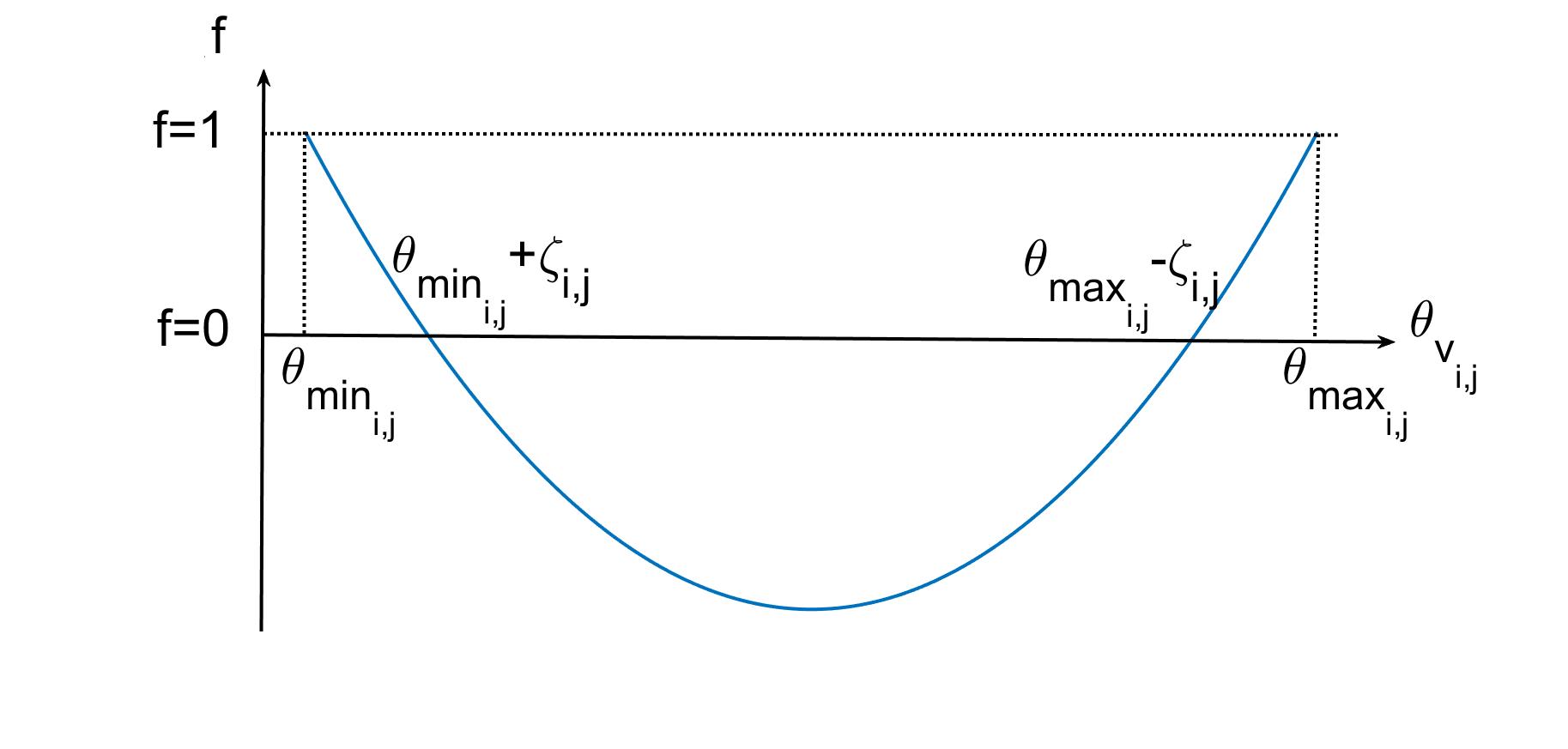}    
		\caption{Convex function $ f(\theta_{v_{i,j}}) $.} 
		\label{fig:convex}
	\end{center}
\end{figure}

The projection operator, denoted as $ \text{Proj} $, for two vectors $ \theta $ and $ y $ is defined as \cite{LavGib11} 
\begin{align} \label{eq:17nx}
&\text{Proj}(\theta, y)\notag \\ &\equiv \left \{\begin{array}{l}y-\frac{\nabla f(\theta)(\nabla f(\theta))^T}{||\nabla f(\theta)||^2}yf(\theta)\  \ if\ f(\theta)>0 \ \\ \ \ \ \ \ \ \ \ \ \ \ \ \ \ \ \ \ \ \ \ \ \ \ \ \ \ \ \ \ \ \ \ \ \ \ \ \ \ \  \& \ y^T\nabla f(\theta)>0 \\  y \ \ \ \ \ \ \ \ \ \ \ \ \ \ \ \ \ \ \ \ \ \ \ \ \  \ \ \ \  otherwise,\end{array}\right.
\end{align}
where $ f(.) $ is a convex and smooth ($ C^1 $) function, and $\nabla(.):\mathbb{R}\rightarrow \mathbb{R}$ is the gradient operator. 
If ${\theta}_v\in \mathbb{R}^{r\times m}$ and $Y\in \mathbb{R}^{r\times m}$ are matrices, the projection operator is defined as
\begin{equation} \label{eq:16nn}
\text{Proj}\big(\theta_v, Y\big)=\big( \text{Proj}(\theta_{v,1}, Y_1), ..., \text{Proj}(\theta_{v,m}, Y_m) \big),
\end{equation}
where $\theta_{v,j}$ and $Y_j$ are the $j$th columns of $\theta_{v}$ and $Y$, respectively, and $ \text{Proj}(\theta_{v,j}, Y_j) $ is defined using (\ref{eq:17nx}). A particular choice of (\ref{eq:17nx}) is given by
\begin{equation} \label{eq:16n}
\text{Proj}\big(\theta_{v,j}, Y_j\big)=\big( \text{Proj}(\theta_{v_{1,j}}, Y_{1,j}), ..., \text{Proj}(\theta_{v_{m,j}}, Y_{m,j}) \big),
\end{equation}
where $\theta_{v_{i,j}}$ and $Y_{i,j}$ are the $ i $th components of $\theta_{v,j}$ and $Y_j$ respectively, and $ \text{Proj}(\theta_{v_{i,j}} , Y_{i,j}): \mathbb{R}\times \mathbb{R}\rightarrow \mathbb{R} $ is an ``element-wise projection" defined as
\begin{align} \label{eq:17n}
&\text{Proj}(\theta_{v_{i,j}}, Y_{i,j})\notag \\ &\equiv \left \{\begin{array}{l}Y_{i,j}-Y_{i,j}f(\theta_{v_{i,j}})\ \ if\ f(\theta_{v_{i,j}})>0 \\ \ \ \ \ \ \ \ \ \ \ \ \ \ \ \ \ \ \ \ \ \ \ \ \ \ \ \ \ \ \ \  \&  \ Y_{i,j}\left( \frac{df(\theta_{v_{i,j}})}{d\theta_{v_{i,j}}}\right) >0\\  Y_{i,j} \ \ \ \ \ \ \ \ \ \ \ \ \ \ \ \ \  otherwise,\end{array}\right.
\end{align}
where $f(.):\mathbb{R}\rightarrow \mathbb{R}$ is a convex function defined as
\begin{equation} \label{eq:18n}
f(\theta_{v_{i,j}})=\frac{(\theta_{v_{i,j}}-\theta_{min_{i,j}}-\zeta_{i,j})(\theta_{v_{i,j}}-\theta_{max_{i,j}}+\zeta_{i,j})}{(\theta_{max_{i,j}}-\theta_{min_{i,j}}-\zeta_{i,j})\zeta_{i,j}},
\end{equation}
where $ \zeta_{i,j} $ is the projection tolerance of the $i,j$th element of $\theta_v$ that should be chosen as $0<\zeta_{i,j}<0.5(\theta_{max_{i,j}}-\theta_{min_{i,j}})$. Also, $\theta_{max_{i,j}}$ and $\theta_{min_{i,j}}$ are the upper and lower bound of the $ i,j $th element of $\theta_{v}$. These bounds also form the projection boundary (see figure \ref{fig:convex}). In comparison to the projection algorithm in \cite{LavGib11}, the projection algorithm (\ref{eq:17n}) is element-wise and the proposed convex function in (\ref{eq:18n}) considers the cases where $ \theta_{min_{i,j}}\neq -\theta_{max_{i,j}} $. In the convex function (\ref{eq:18n}), $f(\theta_{v_{i,j}})=0$ when $\theta_{v_{i,j}}=\theta_{max_{i,j}}-\zeta_{i,j}$ or $\theta_{v_{i,j}}=\theta_{min_{i,j}}+\zeta_{i,j}$, and $f(\theta_{v_{i,j}})=1$ when $\theta_{v_{i,j}}=\theta_{max_{i,j}}$ or $\theta_{v_{i,j}}=\theta_{min_{i,j}}$.

\noindent
\textbf{Lemma 1.} If $\dot{\theta}_{v_{i,j}}=\text{Proj}(\theta_{v_{i,j}},Y_{i,j})$ with initial conditions $\theta_{v_{i,j}}(0)\in \Omega_{i,j}=\{\theta_{v_{i,j}}\in \mathbb{R}|f(\theta_{v_{i,j}})\leq 1\}$, where $f(\theta_{v_{i,j}}):\mathbb{R}\rightarrow \mathbb{R}$ is a convex function, then $\theta_{v_{i,j}}\in \Omega_{i,j}$ for $\forall t\geq 0$.

\noindent 
\textbf{Proof.} The proof of Lemma 1. can be found in \cite{GibAnnLav15}.
\qed

\noindent
\textbf{Lemma 2.}
For $ \theta_{v_{i,j}}^*\in [\theta_{min_{i,j}}+\zeta_{i,j},\ \theta_{max_{i,j}}-\zeta_{i,j}] $, $ \theta_{v_{i,j}}\in \mathbb{R}^{r\times m} $, $ Y\in \mathbb{R}^{r\times m} $ and the projection algorithm in (\ref{eq:17n}) and (\ref{eq:18n}), the following inequality holds:
\begin{equation}\label{eq:e20}
\begin{array}{ll}
tr\bigg( ({\theta}_v^T-{{\theta}_v^*}^T) \big( -Y + \text{Proj}(\theta_v, Y) \big) \bigg) \leq 0,
\end{array}
\end{equation}
where $ tr(.) $ refers to the trace of a matrix.

\noindent
\textbf{Proof.} Let $ I_{i,j}=1 $ if $ f(\theta_{v_{i,j}})>0 $ and $ Y_{i,j}\left( \frac{df(\theta_{v_{i,j}})}{d\theta_{v_{i,j}}}\right) >0 $, and let $ I_{i,j}=0 $, otherwise. Then,
\begin{equation}\label{eq:e20pf}
\begin{array}{ll}
&tr\bigg( ({\theta}_v^T-{{\theta}_v^*}^T) \big( -Y + \text{Proj}(\theta_v, Y) \big) \bigg)\notag \\
&=\displaystyle \sum_{j=1}^{m} \sum_{i=1}^{r} ({\theta}_{v_{i,j}}-{{\theta}_{v_{i,j}}^*}) \big( -Y_{i,j} + \text{Proj}(\theta_{v_{i,j}}, Y_{i,j})\big)\notag \\
&=\displaystyle \sum_{j=1}^{m} \sum_{i=1}^{r} ({\theta}_{v_{i,j}}-{{\theta}_{v_{i,j}}^*}) \big( -Y_{i,j} + Y_{i,j}-Y_{i,j}f(\theta_{v_{i,j}}))\big)I_{i,j}\notag \\
&=\displaystyle \sum_{j=1}^{m} \sum_{i=1}^{r} ({{\theta}_{v_{i,j}}^*}-{\theta}_{v_{i,j}}) Y_{i,j}f(\theta_{v_{i,j}})I_{i,j}\leq 0.
\end{array}
\end{equation}
\qed


\section{Model Reference Adaptive Control Allocation}

\begin{figure}
	\begin{center}
		\includegraphics[width=8.5cm]{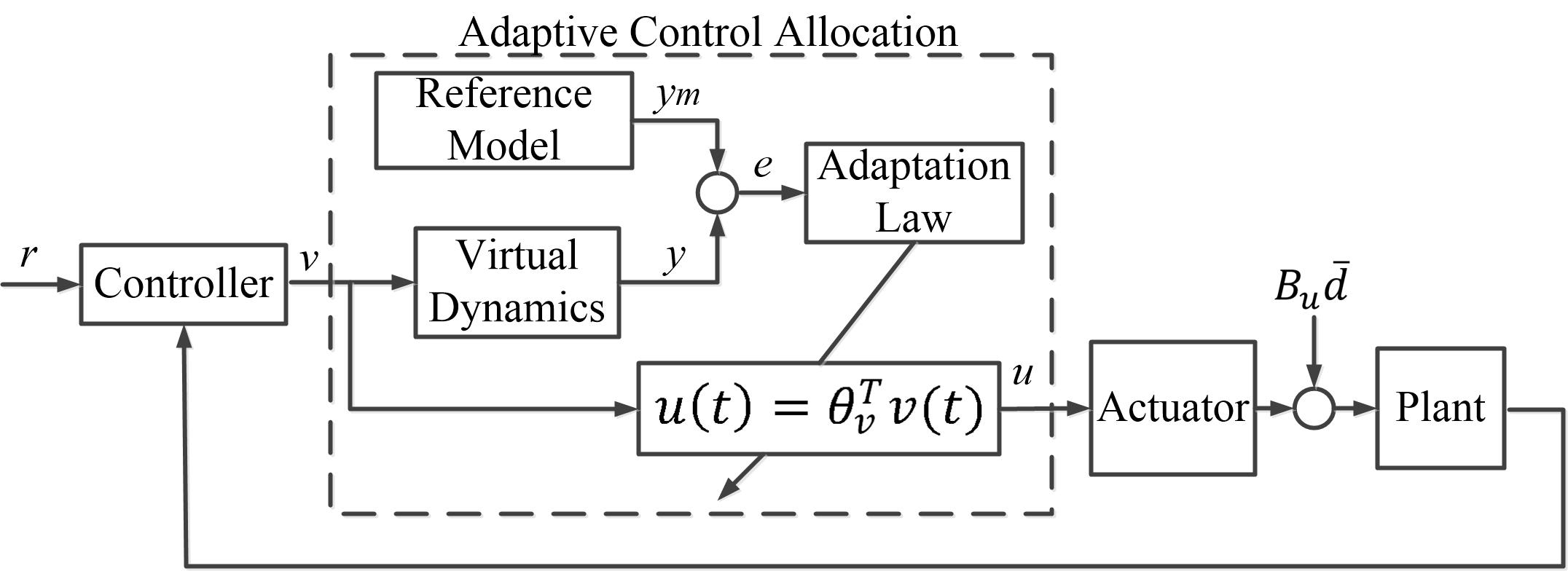}    
		\caption{Block diagram of the closed loop system with the proposed adaptive control allocation method.} 
		\label{fig:block}
	\end{center}
\end{figure}

The closed loop system studied in this paper is presented in Figure \ref{fig:block}. Consider the following plant dynamics,
\begin{equation} \label{eq:e1}
\begin{array}{ll}
\dot{x}&=Ax+B_u(\Lambda u+{d}_u),
\end{array}
\end{equation}
where $x\in \mathbb{R}^{n}$ is the state vector, $u=[u_1,...,u_m]^T\in \mathbb{R}^{m}$ is the actuator input vector, where $u_i\in [-u_{\text{max}_{i}},u_{\text{max}_{i}}]$, $A \in \mathbb{R}^{n\times n}$ is the known state matrix, $B_u \in \mathbb{R}^{n\times m}$ is the known input matrix and $ {d}_u\in \mathbb{R}^m $ is a bounded disturbance input. The matrix $\Lambda \in \mathbb{R}^{m\times m}$ is assumed to be diagonal, with non-negative elements representing actuator effectiveness uncertainty.
It is assumed that the pair $ (A, B_u\Lambda) $ is controllable. 
%
%
Due to actuator redundancy, the input matrix is rank deficient, that is $\text{Rank}(B_u)=r<m$. Consequently, $ B_u $ can be written as $ B_u = B_v B $, where $ B_v \in  \mathbb{R}^{n \times r} $ is a full column rank matrix, i.e. $ \text{Rank}(B_v)=r $, and $ B \in \mathbb{R}^{r \times m} $. The decomposition of $ B_u $ helps exploit the actuator redundancy using control allocation. Employing this decomposition,  (\ref{eq:e1}) can be rewritten as 
\begin{equation} \label{eq:e3new}
\begin{array}{ll}
\dot{x}=Ax+B_v(B\Lambda u+\bar{d}),
\end{array}
\end{equation}
where $ \bar{d}(t) = B{d}_u(t) $ is assumed to have an upper bound $ ||\bar{d}(t)||\leq \bar{L} $, for all $ t\geq 0 $. The control allocation task is to achieve
\begin{equation} \label{eq:e3}
B\Lambda u+\bar{d}= v,
\end{equation}
%
where $ v\in \mathbb{R}^n $ is the virtual control signal and also the output of the nominal controller which will be defined in Section V (See Figure \ref{fig:block}). Considering the following dynamics,
%
\begin{equation} \label{eq:e4}
\dot{y}=A_my+B\Lambda u+\bar{d}-v,
\end{equation}
where $A_m \in \mathbb{R}^{r\times r}$  is a stable (Hurwitz) matrix, a reference model is constructed as
\begin{equation} \label{eq:e5}
\dot{y}_m=A_my_m.
\end{equation}
Defining the actuator input as a mapping from $v$ to $u$,
\begin{equation} \label{eq:e6}
u={\theta}_v^Tv,
\end{equation}
where $\theta_v \in \mathbb{R}^{r\times m}$ represents the adaptive parameter matrix to be determined, and substituting (\ref{eq:e6}) into (\ref{eq:e4}), we obtain
\begin{equation} \label{eq:e7}
\dot{y}=A_my+(B\Lambda{\theta}_v^T-I_r)v+\bar{d},
\end{equation}
where $ I_r $ is an identity matrix of dimension $ r\times r $. 

%
%
It is assumed that there exists an ideal matrix $\theta_v^*$ such that
\begin{equation}\label{eq:e10x}
B \Lambda {\theta_v^*}^T = I_r.
\end{equation}

%
Defining $e=y-y_m$ and subtracting (\ref{eq:e5}) from (\ref{eq:e7}), it follows that
\begin{equation}\label{eq:e11}
\dot{e}=A_me+B\Lambda \tilde{\theta}_v^Tv+\bar{d},
\end{equation}
%
where $ \tilde{\theta}_v=\theta_v-\theta_v^* $.

\noindent
\textbf{Theorem 1.} If the adaptive parameter matrix (\ref{eq:e4}) is updated using the following adaptive law,
\begin{equation}\label{eq:thm1}
\begin{array}{ll}
\dot{\theta}_v=\Gamma_{\theta}\text{Proj}\big(\theta_v, -v{e}^TPB\big),
\end{array}
\end{equation}
where the projection operator ``Proj" is defined in (\ref{eq:17n}), with convex and smooth ($ C^1 $) function $ f(\theta_{v_{i,j}}) $ in (\ref{eq:18n}), and where $ \Gamma_{\theta}=\gamma_{\theta}I_r $, $ \gamma_{\theta}>0 $, then given any initial condition $ e(0)\in \mathbb{R}^r $, $\theta_{v_{i,j}}(0)\in [\theta_{min_{i,j}},\ \ \theta_{max_{i,j}} ]$, and $ \theta_{v_{i,j}}^*\in [\theta_{min_{i,j}}+\zeta_{i,j},\ \ \theta_{max_{i,j}}-\zeta_{i,j}] $, $ e(t) $ and
$ \tilde{\theta}_v(t) $ remain uniformly bounded for all $ t \geq 0 $ and their trajectories converge exponentially to the set
\begin{equation}\label{eq:thm1x}
\begin{array}{ll}
E_1=\{(e,\tilde{\theta}_v):&||e||^2\leq (\dfrac{s\tilde{\theta}_\text{max}^2}{\gamma_{\theta}}+\dfrac{2m^4\bar{L}^2}{\sigma^2})\dfrac{4sm^2}{\sigma},\\
&||\tilde{\theta}_v||\leq \tilde{\theta}_{max} \},
\end{array}
\end{equation}
where constants $ s $, $ \sigma $, $ m $ and $ \tilde{\theta}_{max} $ will be defined in the proof of the theorem.

\noindent
\textbf{Proof.}
Consider a Lyapunov function candidate,
\begin{equation}\label{eq:e14}
V=e^TPe+tr(\tilde{\theta}_v^T\Gamma_{\theta}^{-1}\tilde{\theta}_v\Lambda),
\end{equation}
where $\Gamma_{\theta}=\Gamma_{\theta}^T=\gamma_{\theta} I_r$, $ \gamma_{\theta}>0 $, $tr$ refers to the trace operation and $P$ is the positive definite symmetric matrix solution of the Lyapunov equation,
\begin{equation}\label{eq:e15}
A_m^{T}P + PA_m = -Q,
\end{equation}
and where $Q$ is a symmetric positive definite matrix. The derivative of the Lyapunov function candidate (\ref{eq:e14}) along the trajectories of (\ref{eq:e11})-(\ref{eq:thm1}) can be calculated as
\begin{align}\label{eq:e16}
\dot{V}&={e}^T(A_mP + PA_m)e+2{e}^TPB\Lambda\tilde{\theta}_v^Tv\notag \\
&+2tr(\tilde{\theta}_v^T\Gamma_{\theta}^{-1}\dot{\tilde{\theta}}_v\Lambda)+2e^TP\bar{d} \\
&=-{e}^TQe+2{e}^TPB\Lambda\tilde{\theta}_v^Tv + 2tr(\tilde{\theta}_v^T\Gamma_{\theta}^{-1}\dot{\tilde{\theta}}_v\Lambda)+2e^TP\bar{d}.\notag 
\end{align}
Using the property of the trace operation, $a^Tb=tr(ba^T)$ where $a$ and $b$ are vectors, (\ref{eq:e16}) can be rewritten as
\begin{equation}\label{eq:e17}
\begin{array}{ll}
\dot{V}&=-{e}^TQe+2tr\bigg(\tilde{\theta}_v^T\Big(v{e}^TPB + \Gamma_{\theta}^{-1}\dot{\tilde{\theta}}_v\Big)\Lambda\bigg)+2e^TP\bar{d}.
\end{array}
\end{equation}
Using the following adaptive law,
\begin{equation}\label{eq:e18}
\begin{array}{ll}
\dot{\theta}_v=\Gamma_{\theta}\text{Proj}\big(\theta_v, -v{e}^TPB\big),
\end{array}
\end{equation}
(\ref{eq:e17}) can be written as
\begin{equation}\label{eq:e19}
\begin{array}{ll}
\dot{V}&=-{e}^TQe+2e^TP\bar{d} \\
&+2tr\bigg(\tilde{\theta}_v^T\Big(v{e}^TPB + \text{Proj}\big(\theta_v, -v{e}^TPB\big)\Big)\Lambda\bigg).
\end{array}
\end{equation}

%
%

%
In the absence of a disturbance $ \bar{d} $, it can be shown, by using Lemma 2, that $ \dot{V}\leq 0 $ and therefore $ e $ and $ \tilde{\theta}_v $ are bounded. Furthermore, using Barbalat's lemma it can be shown that the error $ e $ converges to zero. When $ \bar{d} \neq 0 $, it can be shown that all the trajectories converge to a compact set $ E_1 $. To find $ E_1 $, it is necessary to define the following parameters \cite{GibAnnLav15}
\begin{align}
\sigma &\equiv -\displaystyle \max_{i}(\text{Real}({\lambda}_{i}({A}_m))), \label{eq:e26}\\
s &\equiv -\displaystyle \min_{i}({\lambda}_{i}({A}_m+{A}_m^T)/2),\label{eq:e27}\\
a &\equiv ||{A}_m||,\label{eq:e28}
\end{align}

where $ \lambda_i(A_m) $ refers to the $ i $th eigenvalue of the matrix $ A_m $. If the matrix $ Q $ in (\ref{eq:e15}) is selected as an identity matrix of dimension $r\times r$, then the matrix  ${P}$ satisfies the following properties \cite{GibAnnLav15}
\begin{align}
||{P}||\leq \frac{m^2}{\sigma},\label{eq:e29}\\
{\lambda}_\text{min}({P})\geq \frac{1}{2s},\label{eq:e30}
\end{align}

where $ \sigma $ and $s$ are defined in (\ref{eq:e26}) and (\ref{eq:e27}), $ \lambda_\text{min}(.) $ denotes the minimum eigenvalue and $ m=\dfrac{3}{2}(1+4\dfrac{a}{\sigma})^{(r-1)} $, and where $a$ is defined in (\ref{eq:e28}).


Using the Lyapunov function candidate (\ref{eq:e14}), it follows that
\begin{equation}\label{eq:e31}
\begin{array}{ll}
V&=e^T{P}e+tr(\tilde{\theta}_v^T\Gamma_{\theta}^{-1}\tilde{\theta}_v\Lambda) \\
&\leq ||e||^2||{P}||+tr(\tilde{\theta}_v^T\Gamma_{\theta}^{-1}\tilde{\theta}_v\Lambda) \\
&= ||e||^2||{P}||+(1/{\gamma_{\theta}})tr(\tilde{\theta}_v^T\tilde{\theta}_v\Lambda) \\
&\leq ||e||^2||{P}||+(1/{\gamma_{\theta}})||\tilde{\theta}_v||_F^2 \\
&\leq ||e||^2||{P}||+(1/{\gamma_{\theta}})\tilde{\theta}_\text{max}^2,
\end{array}
\end{equation}
where $\Gamma_{\theta}^{-1}=(1/{\gamma_{\theta}})I_r$, $ \gamma_{\theta}>0 $, $\Lambda =diag(\lambda_1, ..., \lambda_m),\ 0< \lambda_i\leq 1$ and considering $ \theta_{v_{i,j}}^*\in [\theta_{min_{i,j}}+\zeta_{i,j},\ \ \theta_{max_{i,j}}-\zeta_{i,j}] $ and using (\ref{eq:18n}), we have $ ||\tilde{\theta}_v(t)||_F\leq \tilde{\theta}_{max} $, for all $ t\geq 0 $, where $ \tilde{\theta}_{max} $ is defined as
\begin{equation} \label{eq:e49x}
\begin{array}{ll}
\tilde{\theta}_{max} \equiv \sqrt{\displaystyle\sum_{i,j}(\theta_{max_{i,j}}-\theta_{min_{i,j}}-\zeta_{i,j})^2}.
\end{array}
\end{equation}

Using (\ref{eq:e31}), we have
\begin{equation}\label{eq:e32}
\frac{V}{||{P}||}-\frac{\tilde{\theta}_\text{max}^2}{\gamma_{\theta} ||{P}||}\leq ||e||^2.
\end{equation}
Since $\theta_{v_{i,j}}^*\in [\theta_{min_{i,j}}+\zeta_{i,j},\ \ \theta_{max_{i,j}}-\zeta_{i,j}] $, using (\ref{eq:e19}), (\ref{eq:e20}), and considering $ Q=I_r $, we have $\dot{V}\leq-||e||^2+2||e||||P\bar{d}||$. In addition, by using the inequality,
\begin{equation}\label{eq:Young}
|xy|\leq \frac{x^2}{2c}+\frac{c|y|^2}{2},
\end{equation}
for $ c=2 $, $ x=||e|| $, $ y=||P\bar{d}|| $, it follows that we have $ 2||e||||P\bar{d}||\leq \frac{1}{2}||e||^2+2||P\bar{d}||^2 $. Recalling that the upper bound of $ \bar{d} $ is $ \bar{L} $, it follows that  $\dot{V}\leq-\frac{1}{2}||e||^2+2||P||^2\bar{L}^2$. Thus, using (\ref{eq:e32}), we have
\begin{align}\label{eq:e33}
\dot{V}(t)&\leq -\frac{1}{2}||e||^2+2||P||^2\bar{L}^2 \notag \\ &\leq -\frac{V}{2||{P}||}+\frac{\tilde{\theta}_\text{max}^2}{2\gamma_{\theta} ||{P}||}+2||P||^2\bar{L}^2\leq -\omega_1 V+\omega_2,
\end{align}
where $\omega_1=\frac{\sigma}{2m^2}$ and $\omega_2=\frac{s}{\gamma_{\theta}}\tilde{\theta}_\text{max}^2+\frac{2m^4\bar{L}^2}{\sigma^2}$.  By using the Gronwall inequality, whose statement is that for $ \dot{V}(t)\leq b(t)V(t)+h(t) $, we have
\begin{align}\label{eq:e33Gron}
{V}(t)&\leq V(0)exp(\int_{\alpha}^{t}b(s)ds)+\int_{\alpha}^{t}h(s)exp(\int_{s}^{t}b(\tau)d\tau)ds, 
\end{align}
(\ref{eq:e33}) can be rewritten as
\begin{equation}\label{eq:e34}
V(t)\leq \big( V(0)-\frac{\omega_2}{\omega_1}\big)e^{-\omega_1t}+\frac{\omega_2}{\omega_1}.
\end{equation}
Using  $e(t)^T{P}e(t)\leq V(t)\leq \big( V(0)-\frac{\omega_2}{\omega_1}\big)e^{-\omega_1t}+\frac{\omega_2}{\omega_1}$ and taking the limits of the leftmost and rightmost  sides as $ t $ goes to infinity, we have
\begin{equation}\label{eq:e35}
\limsup_{t\rightarrow \infty}e(t)^T{P}e(t)\leq \frac{\omega_2}{\omega_1}=(\frac{s\tilde{\theta}_\text{max}^2}{\gamma_{\theta}}+\frac{2m^4\bar{L}^2}{\sigma^2})\frac{2m^2}{\sigma}. 
\end{equation}
By using the following inequality
\begin{equation}\label{eq:e36}
\lambda_\text{min}({P})||e||^2\leq e^T{P}e \leq \lambda_\text{max}({P})||e||^2, 
\end{equation}
and (\ref{eq:e30}), we have
\begin{equation}\label{eq:e37}
\frac{1}{2s}||e||^2\leq \lambda_\text{min}({P})||e||^2 \leq  e^T{P}e.
\end{equation}
By using (\ref{eq:e35}) and taking the limit of both sides of (\ref{eq:e37}) as $ t $ goes to infinity,
\begin{equation}\label{eq:e38}
\limsup_{t\rightarrow \infty}||e(t)||^2 \leq (\dfrac{s\tilde{\theta}_\text{max}^2}{\gamma_{\theta}}+\dfrac{2m^4\bar{L}^2}{\sigma^2})\dfrac{4sm^2}{\sigma}. 
\end{equation}
Therefore, for the initial conditions $e(0)$ and ${\theta}_{v_{i,j}}(0)\in [\theta_{min_{i,j}},\ \ \theta_{max_{i,j}}]$, $e(t)$ and $\tilde{\theta}_v(t)$ are uniformly bounded for all $ t\geq 0$ and system trajectories converge to the following compact set
\begin{equation}\label{eq:e39}
\begin{array}{ll}
E_1=\{(e,\tilde{\theta}_v):&||e||^2\leq (\dfrac{s\tilde{\theta}_\text{max}^2}{\gamma_{\theta}}+\dfrac{2m^4\bar{L}^2}{\sigma^2})\dfrac{4sm^2}{\sigma},\\
&||\tilde{\theta}_v||\leq \tilde{\theta}_{max} \}.
\end{array}
\end{equation}
\qed

It is noted that the bound on $ \tilde{\theta}_{v} $ in (\ref{eq:e39}) is a direct result of Lemma 1 and the definition given in (\ref{eq:e49x}).

The analysis provided above shows that $ \theta_v $ and $ e $ are bounded. Assuming that $v$ is bounded, (\ref{eq:e6}) implies that $u$ is bounded. In the sequel, the boundedness of $ v $ will be established by using a soft saturation bound on  $ v $ during the design of the controller, the effect of which will be analyzed in section \ref{saturation}. Since $A_m$ is Hurwitz, the variable $ y $, whose dynamics is given in (\ref{eq:e4}), is also bounded. Therefore, all the signals in the adaptive control allocation system are bounded.

\vspace*{3mm}
\noindent
\textbf{Remark 1.} Note that $ {\theta}_v^*\in \mathbb{R}^{r\times m} $ is the ideal parameter matrix that should satisfy (\ref{eq:e10x}). Since $ \Lambda $ is unknown, $ {\theta}_v^* $ is also unknown. However, although the diagonal matrix $ \Lambda $ is unknown, the range of its elements can be taken as $ (0,\ 1] $, assuming that the uncertainty originates from possible loss of actuator effectiveness. Thus, using (\ref{eq:e10x}), the range of $ \theta_v^* $ can be obtained, and expressed as $ \theta^*_{i,j}\in [ \theta^*_{{min}_{i,j}},\  \theta^*_{{max}_{i,j}} ]  $. 

\noindent
\textbf{Remark 2.}
Let $ [\theta_{min_{i,j}},\ \theta_{max_{i,j}}]\subset [\theta^*_{{min}_{i,j}},\  \theta^*_{{max}_{i,j}}] $ and $ \theta_{v_{i,j}}^*\notin [\theta_{min_{i,j}}+\zeta_{i,j},\ \ \theta_{max_{i,j}}-\zeta_{i,j}] $, and consider the projection algorithm (\ref{eq:17n}) with convex function (\ref{eq:18n}). Then, $ \tilde{\theta}_{max} $, which was defined in (\ref{eq:e49x}), should be redefined so that $ ||\tilde{\theta}_v(t)||_F\leq \tilde{\theta}_{{MAX}} $, for all $ t\geq 0 $, where $ \tilde{\theta}_{{MAX}} $ is defined as
\begin{equation} \label{eq:e49xx}
\begin{array}{ll}
\tilde{\theta}_{{MAX}}\equiv\\  \sqrt{\displaystyle\sum_{i,j}(\max (|\theta^*_{{max}_{i,j}}-\theta_{min_{i,j}}|,|\theta^*_{{min}_{i,j}}-\theta_{max_{i,j}}|))^2}.
\end{array}
\end{equation}

We note that to delineate two different cases (the ideal parameter $ \theta_v^* $ being inside or outside the projection bounds), the maximum value of adaptive parameter deviation from its ideal value is designated by $ \tilde{\theta}_{max} $ for the former and $ \tilde{\theta}_{MAX} $ for the latter case. Below, we provide a lemma and a theorem, regarding the stability of the control allocation for the latter case, i.e. when $ \theta_{v_{i,j}}^*\notin [\theta_{min_{i,j}}+\zeta_{i,j},\ \ \theta_{max_{i,j}}-\zeta_{i,j}] $.

\noindent
\textbf{Lemma 3.}
For $ \theta_{v_{i,j}}^*\notin [\theta_{min_{i,j}}+\zeta_{i,j},\ \ \theta_{max_{i,j}}-\zeta_{i,j}] $, $ \theta_{v_{i,j}}\in \mathbb{R}^{r\times m} $, $ Y\in \mathbb{R}^{r\times m} $ with $ r\leq m $ and the projection algorithm (\ref{eq:17n})-(\ref{eq:18n}), the following inequality holds:
\begin{equation}\label{eq:e20x}
\begin{array}{ll}
&tr\bigg( ({\theta}_v^T-{{\theta}_v^*}^T) \big( -Y + \text{Proj}(\theta_v, Y) \big) \bigg) \leq \sqrt{r}\tilde{\theta}_{MAX}||Y||.
\end{array}
\end{equation}
\noindent
\textbf{Proof.} For both cases in projection algorithm (\ref{eq:17n}), we have
\begin{equation}\label{eq:proof}
\begin{array}{ll}
&tr\bigg( ({\theta}_v^T-{{\theta}_v^*}^T) \big( -Y + \text{Proj}(\theta_v, Y) \big) \bigg) \\ 
&=\displaystyle \sum_{j=1}^{m} \sum_{i=1}^{r} ({\theta}_{v_{i,j}}-{{\theta}_{v_{i,j}}^*}) \big( -Y_{i,j} + \text{Proj}(\theta_{v_{i,j}}, Y_{i,j})\big)\notag \\
&\leq \displaystyle\sum_{j=1}^{m}\sum_{i=1}^{r}|({\theta}_{v_{i,j}}-{{\theta}_{v_{i,j}}^*})Y_{i,j}f(\theta_{v_{i,j}})|\\
&\leq \displaystyle\sum_{j=1}^{m}\sum_{i=1}^{r}|\tilde{\theta}_{i,j}Y_{i,j}|=tr(|\tilde{\theta}_v^T||Y|)\\
&\leq ||\tilde{\theta}_v||_F||Y||_F\leq \sqrt{r}||\tilde{\theta}_v||_F||Y||\leq \sqrt{r}\tilde{\theta}_{MAX}||Y||,
\end{array}
\end{equation}
where we used the property, $ ||Y||_F\leq \sqrt{min(r,m)}||Y|| $, and $ |\tilde{\theta}_v^T| $ and $ |Y| $ which are the matrices of absolute values of the elements of $ \tilde{\theta}_v^T $ and $ Y $, respectively.
\qed

\noindent
\textbf{Theorem 2.} Consider (\ref{eq:e4}), the reference model (\ref{eq:e5}), the controller (\ref{eq:e6}), and the adaptive law,
\begin{equation}\label{eq:thm2}
\begin{array}{ll}
\dot{\theta}_v=\Gamma_{\theta}\text{Proj}\big(\theta_v, -v{e}^TPB\big),
\end{array}
\end{equation}
where $\Gamma_{\theta}^{-1}=(1/{\gamma_{\theta}})I_r$, $ \gamma_{\theta}>0 $, and the projection is defined in (\ref{eq:17n}) and (\ref{eq:18n}). Assume $ ||v(t)||\leq M $ and $ ||\bar{d}(t)||\leq \bar{L} $ for all $ t\geq 0 $. Then, for any initial condition $ e(0)\in \mathbb{R}^r $, $\theta_{v_{i,j}}(0)\in [\theta_{min_{i,j}},\ \ \theta_{max_{i,j}} ]$, and $ \theta_{v_{i,j}}^*\notin [\theta_{min_{i,j}}+\zeta_{i,j},\ \ \theta_{max_{i,j}}-\zeta_{i,j}] $, $ e(t) $ and
$ \tilde{\theta}_v(t) $ are uniformly bounded for all $ t \geq 0 $ and their trajectories converge exponentially to
\begin{align}\label{eq:thm2x}
\hat{E}_1&=\{(e,\tilde{\theta}_v): ||\tilde{\theta}_v||\leq \tilde{\theta}_{MAX},\ ||e||^2\leq (\dfrac{s\tilde{\theta}_\text{MAX}^2}{\gamma_{\theta}}\notag \\
&+\dfrac{4m^4\bar{L}^2+4r\tilde{\theta}_\text{MAX}^2m^4||B||^2M^2}{\sigma^2})\dfrac{4sm^2}{\sigma} \},
\end{align}
where the constants $ \sigma $, $ s $ and $ m $ are defined in the proof of Theorem 1 and $ \tilde{\theta}_{MAX} $ is defined in Remark 2.

\noindent
\textbf{Proof.} By using (\ref{eq:e14}), (\ref{eq:e15}) with $ Q=I_r $, (\ref{eq:e19}) and (\ref{eq:e20x}) with $ Y=-ve^TPB $, we have 
\begin{align}\label{eq:e33x}
\dot{V}&\leq-||e||^2+2||e||||P||\bar{L}+2\sqrt{r}\tilde{\theta}_{MAX}||Y||\notag \\
&\leq -||e||^2+2||e||||P||\bar{L}+2\sqrt{r}\tilde{\theta}_{MAX}||e||||P||||B||M,
\end{align}
where $ ||\bar{d}||\leq \bar{L} $ and $ v $ is the control
command vector produced by the controller with an upper bound $ M $.

By using the inequality (\ref{eq:Young}) with $ x=||e|| $, $ y=2||P||\bar{L} $, $ c=2 $ for $ 2||e||||P||\bar{L} $ in (\ref{eq:e33x}), and $ x=||e|| $, $ y=2\sqrt{r}\tilde{\theta}_{MAX}||P||||B||M $, $ c=2 $ for $ 2\sqrt{r}\tilde{\theta}_{MAX}||e||||P||||B||M $ in (\ref{eq:e33x}), we obtain that
\begin{align}\label{eq:e33xx}
\dot{V}\leq-\frac{1}{2}||e||^2+4||P||^2\bar{L}^2+4r\tilde{\theta}_{MAX}^2||P||^2||B||^2M^2.
\end{align}
Using (\ref{eq:e32}), we obtain that
\begin{align}\label{eq:e33xxx}
\dot{V}&\leq -\frac{V}{2||{P}||}+\frac{\tilde{\theta}_\text{MAX}^2}{2\gamma_{\theta} ||{P}||}+4||P||^2\bar{L}^2\\ \notag &+4r\tilde{\theta}_\text{MAX}^2||P||^2||B||^2M^2\leq -\hat{\omega}_1 V+\hat{\omega}_2,
\end{align}
where $\hat{\omega}_1=\dfrac{\sigma}{2m^2}$ and $\hat{\omega}_2=\dfrac{s}{\gamma_{\theta}}\tilde{\theta}_\text{MAX}^2+4\dfrac{m^4\bar{L}^2}{\sigma^2}+4\frac{r\tilde{\theta}_\text{MAX}^2m^4||B||^2M^2}{\sigma^2}$, and where $ \sigma $ and $ s $ are defined in (\ref{eq:e26}) and (\ref{eq:e27}). Following the same procedure as for $ E_1 $, $ \hat{E}_1 $ is obtained as

\begin{align}\label{eq:e39x}
\hat{E}_1&=\{(e,\tilde{\theta}_v): ||\tilde{\theta}_v||\leq \tilde{\theta}_{MAX},\ ||e||^2\leq (\dfrac{s\tilde{\theta}_\text{MAX}^2}{\gamma_{\theta}}\notag \\
&+\dfrac{4m^4\bar{L}^2+4r\tilde{\theta}_\text{MAX}^2m^4||B||^2M^2}{\sigma^2})\dfrac{4sm^2}{\sigma} \}.
\end{align}
\qed

\noindent
\textbf{Remark 3.} A discussion about putting an upper bound $ M $ on the control command, without assuming a stable control allocation, is given in section IV.
\vspace*{3mm} 

To obtain fast convergence without introducing excessive oscillations, the open loop reference model (\ref{eq:e5}) is modified to obtain the following closed loop reference model \cite{GibAnnLav15}. 
\begin{equation}\label{eq:e21}
\dot{y}_m=A_my_m-L (y-y_m),
\end{equation}
where $A_m\in \mathbb{R}^{r\times r}$ is Hurwitz, $L=-\ell I_r, \ell >0$, and $I_r\in \mathbb{R}^{r\times r}$ is an identity matrix. Defining $\bar{A}_m=A_m+L$, and subtracting (\ref{eq:e21}) from (\ref{eq:e7}), it follows that
\begin{equation}\label{eq:e22}
\dot{e}=\bar{A}_me+B\Lambda \tilde{\theta}_v^Tv+\bar{d}.
\end{equation}
We assume that the matrix $ \bar{A}_m $ is made Hurwitz through an appropriate selection of $ L $.


\noindent
\textbf{Theorem 3.} Consider (\ref{eq:e4}), the reference model (\ref{eq:e21}), the controller (\ref{eq:e6}), and the adaptive law
\begin{equation}\label{eq:thm3}
\begin{array}{ll}
\dot{\theta}_v=\Gamma_{\theta}\text{Proj}\big(\theta_v, -v{e}^TPB\big),
\end{array}
\end{equation}
where $\Gamma_{\theta}^{-1}=(1/{\gamma_{\theta}})I_r$, $ \gamma_{\theta}>0 $ and the projection is defined by (\ref{eq:17n}) and (\ref{eq:18n}). For any initial condition $ e(0)\in \mathbb{R}^r $, and $\theta_{v_{i,j}}(0)\in [\theta_{min_{i,j}},\ \ \theta_{max_{i,j}} ]$, $ e(t) $ and
$ \tilde{\theta}(t) $ are uniformly bounded for all $ t \geq 0 $ and their trajectories converge exponentially to a closed and bounded set defined either by (\ref{eq:e39xx}) or (\ref{eq:e39xxxx}) in the proof of Theorem 3.

\noindent
\textbf{Proof.} Consider the following Lyapunov function candidate,
\begin{equation}\label{eq:e24x}
V_1=e^T\bar{P}e+tr(\tilde{\theta}_v^T\Gamma_{\theta}^{-1}\tilde{\theta}_v\Lambda),
\end{equation}
where $ \bar{P} $ is the symmetric positive definite matrix solution of the following Lyapunov equation,
\begin{equation}\label{eq:e25}
\bar{A}_m^{T}\bar{P} + \bar{P}\bar{A}_m = -I_r,
\end{equation}
where $ I_r $ is an identity matrix of dimension $ r\times r $. The time derivative of $ V_1 $ along the trajectories of (\ref{eq:e22})-(\ref{eq:thm3}) can be obtained as
\begin{equation}\label{eq:e25x}
\begin{array}{ll}
\dot{V}_1&=-{e}^T\bar{Q}e+2e^TP\bar{d}+2tr\bigg(\tilde{\theta}_v^T\Big(v{e}^T\bar{P}B + \Gamma_{\theta}^{-1}\dot{\tilde{\theta}}_v\Big)\Lambda\bigg).
\end{array}
\end{equation}
%
Aassume first that $\theta_{v_{i,j}}^*\in [\theta_{min_{i,j}}+\zeta_{i,j},\ \ \theta_{max_{i,j}}-\zeta_{i,j}] $.
To find the set to which $e$ and $\tilde{\theta}_v$ converge, it is necessary to define the following parameters \cite{GibAnnLav15}
%
\begin{align}
\bar{\sigma} &\equiv -\displaystyle \max_{i}(\text{Real}({\lambda}_{i}(\bar{A}_m))),\label{eq:e26x}\\
\bar{s} &\equiv -\displaystyle \min_{i}({\lambda}_{i}(\bar{A}_m+\bar{A}_m^T)/2),\label{eq:e27x}\\
\bar{a} &\equiv ||\bar{A}_m||.\label{eq:e28x}
\end{align}

Then, $\bar{P}$ satisfies the following properties \cite{GibAnnLav15}:
\begin{align}
||\bar{P}||\leq \frac{\bar{m}^2}{\bar{\sigma}+2\ell},\label{eq:e29x}\\
{\lambda}_\text{min}(\bar{P})\geq \frac{1}{2(\bar{s}+\ell)},\label{eq:e30x}
\end{align}
where $ \lambda_\text{min}(.) $ denotes the minimum eigenvalue and $ \bar{m}=\dfrac{3}{2}(1+4\dfrac{\bar{a}}{\bar{\sigma}})^{(r-1)} $.


Proceeding as in the proof of Theorem 1, and using (\ref{eq:e26x}-\ref{eq:e28x}), for the initial conditions $e(0)$ and $||\tilde{\theta}_v(0)||\leq \tilde{\theta}_{max}$, $e$ and $\tilde{\theta}_v$ can be shown to be uniformly bounded and converge to the following set,
\begin{align}\label{eq:e39xx}
E_2=&\{(e,\tilde{\theta}_v):||e||^2\leq (\dfrac{(\bar{s}+\ell)\tilde{\theta}_\text{max}^2}{\gamma_{\theta}}+\dfrac{2\bar{m}^4\bar{L}^2}{(\bar{\sigma}+2\ell)^2})\notag \\  &\times\dfrac{4(\bar{s}+\ell)\bar{m}^2}{(\bar{\sigma}+2\ell)}, ||\tilde{\theta}_v||\leq \tilde{\theta}_{max} \}.
\end{align}

If $ \theta_{v_{i,j}}^*\notin [\theta_{min_{i,j}}+\zeta_{i,j},\ \ \theta_{max_{i,j}}-\zeta_{i,j}] $, proceeding similar as in the proof of Theorem 2, the convergence set is characterized as
\begin{align}\label{eq:e39xxxx}
\hat{E}_2=&\{(e,\tilde{\theta}_v):||e||^2\leq (\dfrac{(\bar{s}+\ell)\tilde{\theta}_\text{MAX}^2}{\gamma_{\theta}}+\dfrac{4\bar{m}^4\bar{L}^2}{(\bar{\sigma}+2\ell)^2}\notag \\&+\dfrac{4r\tilde{\theta}_\text{MAX}^2\bar{m}^4||B||^2M^2}{(\bar{\sigma}+2\ell)^2})\dfrac{4(\bar{s}+\ell)\bar{m}^2}{(\bar{\sigma}+2\ell)}, \notag \\
&||\tilde{\theta}_v||\leq \tilde{\theta}_{MAX} \}.
\end{align}
\qed

\addtolength{\textheight}{-3cm}   


\section{Determination of the projection boundaries}\label{project}

In the previous section, the adaptive control allocator was designed based on the projection operator and proved to be stable. In this section, the selection of the projection boundaries, which define the bounds on adaptive control parameters, is explained. The projection boundaries are determined to satisfy two requirements: 1) The actuator command signals should not saturate the actuators and 2) a specific requirement in the controller design, which will be provided in the following subsections, to obtain a stable closed loop system, including the controller, the control allocator and the plant (see Fig. \ref{fig:block}), should be satisfied. The design procedure to achieve these goals is composed of three main steps. In the first step, an attainable set for virtual control signal vector $ v $ is found in the absence of disturbance based on the actuators constraints and $ v=B\Lambda u $. In the second step, using the calculated attainable set for $ v $, projection bounds are calculated to satisfy $ -u_{max}\leq \theta_v^Tv\leq u_{max} $. In the first two steps, the attainable sets are obtained, and as long as the signals are inside these set, we can guarantee that the actuator constraints are satisfied. In the third step, a subset of the projection boundaries calculated in step 2 that satisfies an overall closed loop stability requirement is determined.

\textbf{Step 1}

In this step, realizable values of virtual control signals are found.

Note that the actuator constraints are known: $ u(t)\in \Omega_u $, where $ \Omega_u=\{[u_1, ..., u_m]^T:-u_{max_{i}}\leq u_i\leq u_{max_{i}},i=1, ..., m\} $. Therefore, using $ \Omega_u $, the set $ \Omega_v\in \mathbb{R}^{r} $, defining all realizable values of the virtual control input $ v $, can be obtained as $ \Omega_v=\{v:v=Bu, u\in \Omega_{u}, B^{\dagger}v\in \Omega_u\} $. Note that $ \Omega_v $ also defines the upper and lower bounds of each element of the realizable virtual control, $ v=[v_1, ..., v_r]^T $. To make sure that $ v_i $ remains within its realizable bounds, $ v_i\in[-M_i, M_i] $, $ \forall i=1, ..., r $, we use a soft saturation bound on the control signal $ v $. In Section~V, we will design the controller by taking this saturation bound into account.

\textbf{Step 2}

The projection boundaries that limit the adaptive parameters are calculated in this step, to make sure that the actuator signal vector $ u $ does not saturate the actuators.

From Step 1, we obtained the attainable set for the virtual control signals ($ \Omega_v $). The actuator limits, $ u_{max} $ and $ u_{min}=-u_{max} $ are known. With this information, the set $ \Omega_{\theta}=\{vec(\theta_v): -u_{max}\leq \theta_v^Tv \leq u_{max}, v\in \Omega_v \} $, where $ \text{vec}(.):\mathbb{R}^{r\times m}\rightarrow \mathbb{R}^{rm} $ puts the elements of a matrix in a column vector, can be obtained. Note that $ \theta_I^*\in \Omega_{\theta} $, that is, in the absence of any uncertainty, the ideal adaptive parameter matrix, $ \theta_I^* $, always exists in $ \Omega_{\theta} $. This leads to the smallest convergence set for error trajectories when $ \Lambda=I $. 

\textbf{Step 3}

In this step, a subset of $ \Omega_{\theta} $, which satisfies a necessary condition for controller stability, is obtained. This subset of $ \Omega_{\theta} $ also determines the ultimate projection boundaries, and is denoted by $ \Omega_{\text{proj}} $.

After Step 2, establishing that the control allocation output, which is the actuator input signal vector $ u $, does not saturate the actuators, the plant dynamics (\ref{eq:e3new}) can be rewritten, by using (\ref{eq:e6}), (\ref{eq:e10x}) and defining $ \tilde{\theta}_v=\theta_v-\theta_v^* $, as
\begin{equation}\label{eq:e43x}
\begin{aligned}
\dot{x}&=Ax+B_v(B\Lambda u+\bar{d}) \\
&=Ax+B_v(B\Lambda \theta_v^Tv+\bar{d}) \\
&=Ax+B_v(I+B\Lambda \tilde{\theta}_v^T)v+B_v\bar{d}.
\end{aligned} 
\end{equation}
Defining $ \Delta B \equiv B\Lambda\tilde{\theta}_v^T $, and substituting in (\ref{eq:e43x}), it follows that 
\begin{equation}\label{eq:e43xx}
\begin{aligned}
\dot{x}&=Ax+B_v(v+d),
\end{aligned} 
\end{equation}
where $ d=\Delta Bv+\bar{d} \in \mathbb{R}^r $. 

To be able to design a stabilizing controller, one must make sure that the $ i $th element of the disturbance vector $ d=[d_1, ..., d_r]^T $ in (\ref{eq:e43xx}), is smaller in absolute value than the upper bound of the $ i $th element of the virtual control input which was defined in Step 1, that is $ |d_i|<M_i $, $ i=1, ...,r $.
Since each $ d_i=\text{row}_i(\Delta B)v+\bar{d}_i $, where $ \bar{d}_i $ is the $ i $th element of $ \bar{d} $, and $ \text{row}_i(.) $ designates the $ i $th row of a matrix, satisfying the following condition ensures that $ |d_i|<M_i $, $ i=1, ...,r $:
\begin{equation}\label{eq:e43xxx}
\begin{aligned}
M_i-||\text{row}_i(\Delta B)||M_{\text{max}}>|\bar{d}_i|,\ i=1, ...,r,
\end{aligned} 
\end{equation}
where $ M_{\text{max}}=\max_{i}{M_i} $. A necessary condition for satisfying the inequality (\ref{eq:e43xxx}), is $ ||\text{row}_i(\Delta B)||=||\text{row}_i(B\Lambda\tilde{\theta}_v^T)||<\frac{M_i}{M_{\text{max}}} $ for all $ i=1, ..., r $. (Sufficient conditions required to satisfy (\ref{eq:e43xxx}) will be discussed later in Remark 4).  Thus, the elements of the matrix $ \theta_v $ should be properly bounded in order to satisfy the necessary condition $ ||\text{row}_i(B\Lambda\tilde{\theta}_v^T)||<\frac{M_i}{M_{\text{max}}} $ for all $ i=1, ..., r $, and for all $ \Lambda\in \Omega_{\Lambda_1} $, where $ \Omega_{\Lambda_1} \subset \Omega_{\Lambda} $, and $ \Omega_{\Lambda} $ is the set of all $ m\times m $ diagonal matrices with elements in $ [0,1] $; furthermore, $ \Omega_{\Lambda_1} $ should have diagonal elements $ \lambda_i\in \left( \gamma, 1\right]  $, where $ \gamma $ is precisely defined later in Theorem 4. 


\noindent
\textbf{Remark 4.} 
In order to find a non-empty set, for which, the elements of $ \theta_v $ satisfy the necessary condition discussed above, an optimization problem needs to be solved over the following set,
\begin{align}\label{eq:cvxxxxf}
E=\{&\text{vec}(\theta_v):||\text{row}_i(B\Lambda\theta_v^T-I_r)||^2\leq \frac{M_i^2}{M_{\text{max}}^2}-\epsilon, \notag\\
&\Lambda\in \Omega_{\Lambda_1}, i=1, ..., r \},
\end{align} 
where $ \text{vec}(.):\mathbb{R}^{r\times m}\rightarrow \mathbb{R}^{rm} $ puts the elements of a matrix in a column vector and $ \epsilon $ is a small positive constant used to have a close set, since typical numerical optimizers only optimize over a close set. Note that $ B\Lambda\tilde{\theta}_v^T=B\Lambda(\theta_{v}^T-{\theta_v^*}^T)=B\Lambda\theta_{v}^T-I_r $.

\begin{figure}
	\begin{center}
		\includegraphics[width=6.0cm]{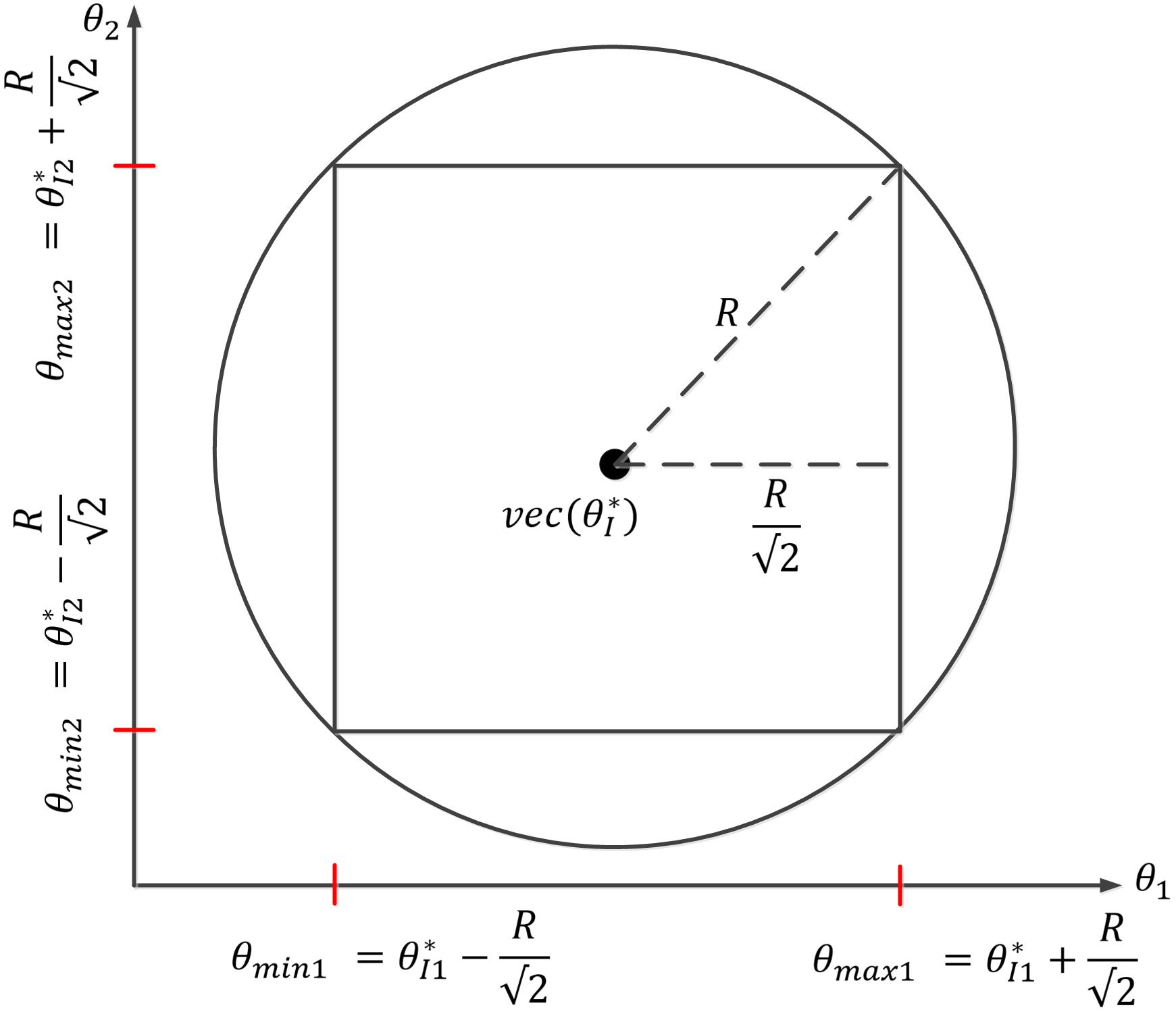}    %
		\caption{Projection boundaries when there are two adaptive parameters. The circle defines the border of the neighborhood obtained from the optimization. The square defines the projection boundaries for $ \theta_1 $ and $ \theta_2 $.} 
		\label{fig:circle}
	\end{center}
\end{figure}

\vspace{0.3cm}
The optimization problem,
\begin{equation}\label{eq:e41_1xxxx}
\begin{array}{ll}
R^2=&\displaystyle \min_{\theta_v} \left(  \text{vec}(\theta_v-\theta_I^*)^T\text{vec}(\theta_v-\theta_I^*)\right) \\
s. t. \ &\left\| \text{row}_i(B\Lambda\theta_v^T-I_r) \right\|^2= \frac{M_i^2}{M_{\text{max}}^2}-\epsilon,\ \ i=1, ..., r,
\\&\Lambda\in \Omega_{\Lambda_1},
\\&\text{vec}(\theta_v)\in \Omega_{\theta},
\end{array}
\end{equation}
which needs to be solved offline, finds the minimum distance, $ R $, from the $ \text{vec}(\theta_I^*) $ to the boundary of the set (\ref{eq:cvxxxxf}). Figure \ref{fig:circle} depicts a visualization of the projection boundaries for the case when there are only two adaptive parameters, $ \theta_1 $ and $ \theta_2 $. 
It is noted that the calculated $ \theta_{max_{i,j}} $ and $ \theta_{min_{i,j}} $ are not unique, and different boundaries can be found by defining different cost functions in (\ref{eq:e41_1xxxx}). 
After calculating $ \theta_{max_{i,j}} $ and $ \theta_{min_{i,j}} $, the projection parameters region is obtained as,
\begin{equation}\label{eq:e43xxy}
\begin{aligned}
\Omega_{\text{proj}}=\{vec(\theta_v): \theta_{i,j}\in [\theta_{max_{i,j}},\ \theta_{max_{i,j}}] \}.
\end{aligned} 
\end{equation}

\noindent
\textbf{Remark 5.}
For all elements of $ \Omega_{\Lambda_1} $, the optimization problem (\ref{eq:e41_1xxxx}) finds the largest neighborhood of $ \theta_{I}^* $ in $ \Omega_{\theta} $ that satisfies the necessary condition. This neighborhood is an n-sphere, with the center at $ \text{vec}(\theta_I^*) $ and with the radius $ R $.

\noindent
\textbf{Remark 6.}
The reason to include $ \theta_{I_{i,j}}^* $ inside the projection boundaries is that it is preferred that $ e $ converges to the smallest set around zero. For this, $  \theta_{I_{i,j}}^*  $ should be inside $ [\theta_{min_{i,j}}+\zeta_{i,j},\ \ \theta_{max_{i,j}}-\zeta_{i,j}] $ (see (\ref{eq:thm1x}) and (\ref{eq:thm2x})).

%
%


%
%
%


In order to show that the optimization problem (\ref{eq:e41_1xxxx}) is feasible, it should be proven that the set $ E $ always includes $ \text{vec}(\theta_I^*) $.

\noindent
\textbf{Theorem 4.} The set $ \Upsilon= \{\text{vec}(\theta_v):||\text{row}_i(B\Lambda\theta_v^T-I_r)||^2\leq\frac{M_i^2}{M_{\text{max}}^2}-\epsilon, \Lambda\in \Omega_{\Lambda_1}\subset \Omega_{\Lambda}, i=1, ...,r\} \cap \text{vec}(\theta_I^*) \neq \varnothing $ when $ \lambda_{\text{min}}(\Lambda)>\gamma\equiv\max_{i}(1-\sqrt{\frac{\gamma_{M_i}}{\gamma_{B_i}}}) $, where $ \lambda_{\text{min}}(.) $ denotes the minimum eigenvalue, $ \gamma_{B_i}\equiv ||\text{row}_i(B)||||B^T(BB^T)^{-1}||$ and $ \gamma_{M_i}\equiv \frac{M_i^2}{M_{\text{max}}^2}-\epsilon$.

\textbf{Proof.}
To prove the non-emptyness of $ \Upsilon $, we should show that $ ||\text{row}_i(B\Lambda{\theta_I^*}^T-I_r)||^2\leq\frac{M_i^2}{M_{\text{max}}^2}-\epsilon $. Using (\ref{eq:cvxxxxf}) and the definition of $ {\theta_I^*}^T=B^T(BB^T)^{-1} $, we have,
\begin{align}\label{eq:67z}
&||\text{row}_i(B\Lambda{\theta_I^*}^T-I_r)||^2\notag \\
=&||\text{row}_i(B\Lambda B^T(BB^T)^{-1}-I_r)||^2\notag \\
=&||\text{row}_i(B\Lambda B^T(BB^T)^{-1}-BB^T(BB^T)^{-1})||^2\notag \\
=&||\text{row}_i(B(\Lambda-I_m)B^T(BB^T)^{-1})||^2\notag\\
\leq&||\text{row}_i(B(\Lambda-I_m))||^2||B^T(BB^T)^{-1}||^2\notag \\
\leq&(\lambda_{\text{max}}(\Lambda-I_m))^2||\text{row}_i(B)||^2||B^T(BB^T)^{-1}||^2\notag \\
=&(\lambda_{\text{min}}(\Lambda)-1)^2||\text{row}_i(B)||^2||B^T(BB^T)^{-1}||^2,
\end{align}
where $ \lambda_{\text{min}}(.) $ denotes the minimum eigenvalue. It is noted that since $ \Lambda $ and $ I_m $ are diagonal matrices and diagonal elements of $ \Lambda $ are between zero and one, we have, $ (\lambda_{\text{max}}(\Lambda-I_m))^2=(\lambda_{\text{min}}(\Lambda)-1)^2 $. Therefore, in order to show that $ ||\text{row}_i(B\Lambda{\theta_I^*}^T-I_r)||^2\leq\frac{M_i^2}{M_{\text{max}}^2}-\epsilon $, for all $ i=1, ..., r $, we should satisfy
\begin{align}\label{eq:67zz}
&(\lambda_{\text{min}}(\Lambda)-1)^2||\text{row}_i(B)||^2||B^T(BB^T)^{-1}||^2\leq\frac{M_i^2}{M_{\text{max}}^2}-\epsilon\notag \\
&\Rightarrow -\sqrt{\frac{\gamma_{M_i}}{\gamma_{B_i}}}\leq\lambda_{\text{min}}(\Lambda)-1\leq\sqrt{\frac{\gamma_{M_i}}{\gamma_{B_i}}},\ i=1, ..., r,
\end{align}
where $ \gamma_{B_i}\equiv ||\text{row}_i(B)||||B^T(BB^T)^{-1}||$ and $ \gamma_{M_i}=\frac{M_i^2}{M_{\text{max}}^2}-\epsilon $ for all $ i=1, ..., r $. Since the maximum value for the diagonal elements of $ \Lambda $ is one, the only condition that should be satisfied is that $1-\sqrt{\frac{\gamma_{M_i}}{\gamma_{B_i}}}\leq\lambda_{\text{min}}(\Lambda) $ for $ i=1, ..., r $ or
$ \gamma\equiv\max_{i}(1-\sqrt{\frac{\gamma_{M_i}}{\gamma_{B_i}}})\leq\lambda_{\text{min}}(\Lambda) $.
\qed

Based on the above theorem, and by defining $ \gamma_{B_i}\equiv ||\text{row}_i(B)||||B^T(BB^T)^{-1}||$, and $ \gamma\equiv \max_{i}(1-\frac{M_i}{\gamma_{B_i}M_{\text{max}}})$, the definition of $ \Omega_{\Lambda} $ should be modified as
\begin{align}\label{eq:lamda}
\Omega_{\Lambda_1}=\{\Lambda:\Lambda\in\mathbb{D}^{m\times m},\ \text{diag}_i(\Lambda)\in (\gamma, 1],\ i=1, ..., m  \},
\end{align}
where $ \mathbb{D} $ denotes the set of real diagonal matrices, and $ \text{diag}_i(.):\mathbb{R}^{m\times m}\rightarrow \mathbb{R} $ provides the $ i $th diagonal element of square matrices.

\noindent
\textbf{Remark 7.} Using $ \Omega_{\Lambda_1} $, defined in (\ref{eq:lamda}), and $ \Omega_{\text{proj}} $, defined in step 3, the upper bound on $ ||\text{row}_i(\Delta B)||=||\text{row}_i(B\Lambda \tilde{\theta}_v^T)||$ can be found, which we denote as $ \bar{\rho}_i $. Therefore, recalling that $ M_i $ is defined as the upper bound on the absolute value of $ i $th virtual control signal $ v_i $, the disturbance $ \bar{d}_i $, which is defined after (8), should be smaller than $ M_i-\bar{\rho}_iM_{\text{max}} $, i.e. $ \bar{d}_i< M_i-\bar{\rho}_iM_{\text{max}} $. Note that $ \bar{\rho}_i<\frac{M_i}{M_{\text{max}}} $ is guaranteed by the solution of (\ref{eq:e41_1xxxx}). Therefore, the condition $ |d_i|< M_i $ is satisfied.

\section{Controller design}
In this section, a design procedure for the controller that generates the virtual control signal $ v $ in (\ref{eq:e43xx}) is proposed.

During the design of the controller, the following two assumptions are made about the plant dynamics:

\noindent
\textbf{Assumption 1.} The dynamics in (\ref{eq:e43xx}) can be written as 
\begin{equation}\label{eq:assump1}
\begin{bmatrix}
\dot{x}^{(1)}\\ \dot{x}^{(2)}
\end{bmatrix} = \begin{bmatrix}
A_{1,1} & A_{1,2}\\ A_{2,1} & A_{2,2}
\end{bmatrix}\begin{bmatrix}
x^{(1)} \\ x^{(2)}
\end{bmatrix}+B_v(v+d),\ \ \ 
y=C\begin{bmatrix}
x^{(1)} \\ x^{(2)}
\end{bmatrix},
\end{equation}
where ${A}_{1,1}\in \mathbb{R}^{(n-r)\times (n-r)}$ is a Hurwitz matrix, ${A}_{1,2}\in \mathbb{R}^{(n-r)\times r}$, ${A}_{2,1}\in \mathbb{R}^{r\times (n-r)}$, ${A}_{2,2}\in \mathbb{R}^{r\times r}$, ${x}^{(1)}\in \mathbb{R}^{(n-r)}$, ${x}^{(2)}\in \mathbb{R}^r$, $ y\in \mathbb{R}^r $ and $ C=[0_{r\times (n-r)}\ I_r] $.

\noindent
\textbf{Assumption 2.} The matrix $ B_v\in \mathbb{R}^{n\times r} $ is in the form $ [0_{r\times (n-r)}\ I_r]^T $. 

Both of the above assumptions are justified for typical aircraft models \cite{Duc09, SteLew15}. In the simulation section, these assumptions are validated using the AeroData Model in Research Environment (ADMIRE) \cite{Har05, Yil11a}. 

\noindent
\textbf{Remark 8.} For systems for which Assumption 2 does not hold, given that $ B_v $ has full column rank, it is possible to find (see \cite{CorCri10, Ant06}) a transformation matrix, $ T_B $, such that $ \hat{B}_v=T_BB_v=[0_{r\times (n-r)}\ I_r]^T $. However, employing this transformation may lead to a state space realization which violates Assumption1.


\noindent
\textbf{Remark 9.} It is desired to design a controller which makes $ y=Cx=x^{(2)} $ follow the reference input. Since $ A_{1,1} $ is Hurwitz, by Assumption 1, showing that the states $ x^{(2)} $ are bounded will be sufficient to demonstrate for the boundedness of $ x^{(1)} $.

\subsection{\textbf{{Dynamics on the Time Varying Sliding Surface}}}

The sliding surface, inspired by \cite{CorCri10}, is given as
\begin{align}\label{eq:e52}
&s({x}^{(2)}(t),{x}^{(2)}(t_0),t)=\notag \\
&{x}^{(2)}(t)-{x}^{(2)}(t_0)e^{-\bar{\lambda}(t-t_0)}-\frac{2}{\pi}{r}(t)\text{tan}^{-1}\big(\bar{\lambda}(t-t_0)\big)=0,
\end{align}
where $ \bar{\lambda}>0 $ is a scalar parameter,  ${x}^{(2)}\in \mathbb{R}^r$ is defined in (\ref{eq:assump1}), $ s\in \mathbb{R}^{r} $ is the sliding surface, and $ r(t)\in \mathbb{R}^r $ is the reference to be tracked.


The response of a system controlled by a sliding mode controller includes two phases. The first phase is called the reaching phase. During this phase, the controller drives the system towards the sliding surface so that $s(t)\to 0$. In the second, sliding phase, the trajectory
evolves on the sliding manifold. For the sliding surface (\ref{eq:e52}), no reaching phase exists since the sliding surface is a function of the initial condition and the trajectories are on the sliding surface at $ t = t_0 $ i.e. $ s({x}^{(2)}(t),{x}^{(2)}(t_0),t_0) = 0 $. In the next subsection, via Theorem 5, the control law $ v $ that ensures that the trajectories remain on the sliding surface for all $ t\geq t_0 $ is provided.

Using (\ref{eq:e52}), the trajectories of $ x^{(2)} $ on the sliding surface satisfy
\begin{align}\label{eq:e54}
{x}^{(2)}(t)&={x}^{(2)}(t_0)e^{-\bar{\lambda}(t-t_0)}+\frac{2}{\pi}{r}(t)\text{tan}^{-1}\big(\bar{\lambda}(t-t_0)\big).
\end{align}
For $ A $ satisfying Assumption 1, and if (\ref{eq:e54}) holds, it follows that
\begin{align}\label{eq:e55x}
\dot{{x}}^{(1)}&={A}_{1,1}{x}^{(1)}+{A}_{1,2}[{x}^{(2)}(t_0)e^{-\bar{\lambda}(t-t_0)}\notag \\
&+\frac{2}{\pi}{r}(t)\text{tan}^{-1}(\bar{\lambda}(t-t_0))].
\end{align}
By defining $ G_1\equiv {A}_{1,2}x^{(2)}(t_0) $, and $ G_2(t)=\frac{2}{\pi}A_{1,2}{r}(t)$ $\text{tan}^{-1}(\bar{\lambda}(t-t_0)) $, we have,
\begin{align}\label{eq:e56}
\dot{{x}}^{(1)}&=A_{1,1}{x}^{(1)}+G_1e^{-\bar{\lambda}(t-t_0)}+G_2(t)=A_{1,1}{x}^{(1)}+g(t),
\end{align}
where $g(t)\equiv G_1e^{-\bar{\lambda}(t-t_0)}+G_2(t)$. 

\noindent
\textbf{Lemma 4.} When $ x^{(2)}(t) $ is on the sliding surface (\ref{eq:e52}), $ x^{(1)}(t) $ and $ x^{(2)}(t) $ are bounded and for all $ t\geq t_0 $, $ ||x^{(1)}(t)||\leq k\bar{x}^{(1)}(t_0)+K_2\bar{x}^{(2)}(t_0)+K_2\bar{{r}} $, where $ K_2=\frac{k}{\xi}||A_{1,2}|| $, and where $ k $ and $ \xi $ are constants. Also, $ \bar{x}^{(1)}(t_0) $, $ \bar{x}^{(2)}(t_0) $ and $ \bar{{r}} $ are the upper bounds of $ ||{x}^{(1)}(t_0)|| $, $ ||{x}^{(2)}(t_0)|| $ and $ \sup_{t\geq t_0}||r(t)|| $, respectively. Furthermore, $ \lim_{t\rightarrow \infty}y(t)=r(t) $. 

\noindent
\textbf{Proof.}
Per Assumption 1, $ A_{1,1} $ is Hurwitz, hence the homogeneous system $\dot{{x}}_{h}^{(1)}(t)=A_{1,1}{x}_{h}^{(1)}(t)$ is globally exponentially stable at the origin. The solution of this system is given as $ x_h^{(1)}(t)=\Phi(t,t_0)x_h^{(1)}(t_0) $, where $ \Phi(t,t_0) $ is the state transition matrix and there exist constants $ k>0 $ and $ \xi>0 $ such that
\begin{align}\label{eq:e57}
||\Phi(t,t_0)||\leq ke^{-\xi(t-t_0)},\ \ \forall t\geq t_0.
\end{align}

Since the state transition matrices of $ \dot{x}_{h}^{(1)}(t)=A_{1,1} x_{h}^{(1)}(t) $ and $ \dot{x}^{(1)}(t)=A_{1,1} x^{(1)}(t)+g(t) $ are the same, we use the state transition matrix $ \Phi(t,t_0) $ used in (\ref{eq:e57}) to provide the solution of (\ref{eq:e56}) as 
\begin{align}\label{eq:e58}
{x}^{(1)}(t)=\Phi(t,t_0){x}^{(1)}(t_0)+\int_{t_0}^{t}\Phi(t,\eta)g(\eta)d\eta.
\end{align}
Taking the norm of both sides of (\ref{eq:e58}), we obtain that
\begin{align}\label{eq:e59xx}
||{x}^{(1)}(t)||&\leq||\Phi(t,t_0){x}^{(1)}(t_0)||+\int_{t_0}^{t}||\Phi(t,\eta)||||g(t)||d\eta.
\end{align}
Using the definition of $ g(t) $ given after (\ref{eq:e56}), it follows that $ ||g(t)||=||G_1e^{-\bar{\lambda}t}+G_2(t)||\leq ||G_1||+\sup_{t\geq t_0}||G_2(t)|| $. Note that $ G_2(t) $ is a function of the reference input $ r(t) $, and that $ \sup_{t\geq t_0}||G_3(t)|| $ exists. Therefore, $ ||g(t)||\leq ||A_{1,2}||||x^{(2)}(t_0)||+||A_{1,2}||||r(t)|| $. Defining $ K_1=||A_{1,2}||||x^{(2)}(t_0)||+||A_{1,2}||\bar{{r}} $, where $\bar{{r}}$ is the upper bound on $ ||r(t)|| $ for $ t\geq 0 $, and using (\ref{eq:e57}), (\ref{eq:e59xx}) can be written as,
\begin{align}\label{eq:e59x}
||{x}^{(1)}(t)||&\leq k||x^{(1)}(t_0)||e^{-\xi(t-t_0)}+kK_1\int_{t_0}^{t}e^{-\xi(t-\eta)}d\eta\notag\\
&\leq k||x^{(1)}(t_0)||e^{-\xi(t-t_0)}+kK_1\frac{1}{\xi}(1-e^{-\xi(t-t_0)})\notag\\
&\leq k||x^{(1)}(t_0)||+kK_1\frac{1}{\xi}\notag\\
&\leq k\bar{x}^{(1)}(t_0)+K_2\bar{x}^{(2)}(t_0)+K_2\bar{r},
\end{align}
where $ K_2=\frac{k}{\xi}||A_{1,2}|| $, while $ \bar{x}^{(1)}(t_0) $ and $ \bar{x}^{(2)}(t_0) $ represent bounds on $ ||x^{(1)}(t_0)|| $ and $ ||x^{(2)}(t_0)|| $, respectively. Since the reference signal $ r(t) $, $ x^{(1)}(t_0) $ and $ x^{(2)}(t_0) $ are bounded, (\ref{eq:e59x}) shows that $ x^{(1)}(t) $ is bounded. Since $ x(t_0) $ and $ r(t) $ are bounded, it can be shown, using (\ref{eq:e54}), that $ x^{(2)}(t) $ is bounded and converges to $ r(t) $. Since $ y=x^{(2)} $, this completes the proof.
\qed


\subsection{\textbf{{Control Law}}}

We now describe the control law and characterize
its properties.

\noindent
\textbf{Definition 1.} $ \text{sign}_\text{v}(a) $, where $ a $ is a column vector, is a diagonal matrix whose elements are the signs of the elements of the vector $ a $. For example, $ \text{sign}_\text{v}([a_1\ a_2]^T)=\text{diag}(\text{sign}(a_1),\ \text{sign}(a_2)) $, where $ a_1 $ and $ a_2 $ are scalars.

\noindent
\textbf{Definition 2.} $ |a|_\text{v}\equiv \text{sign}_\text{v}(a)a $ and $ |a^T|_\text{v}\equiv a^T\text{sign}_\text{v}(a) $, where $ a $ is a column vector and $ \text{sign}_v(.) $ is defined in Definition 3. For example, $ |[a_1\ a_2]|_\text{v}=[a_1\ a_2]\text{sign}_\text{v}([a_1\ a_2]^T)=[|a_1|\ |a_2|] $, where $ a_1 $ and $ a_2 $ are scalars.

\noindent
\textbf{Theorem 5.} Consider the dynamics in (\ref{eq:assump1}), with disturbance $ {d} $, $ t_0=0 $, and the control law, 
\begin{align}\label{eq:e60}
{{v}}(t)=&-{A}_{2,1}x^{(1)}(t)-A_{2,2}x^{(2)}(t)-\bar{\lambda}x^{(2)}(0)e^{-\bar{\lambda}t}\notag \\
&+\frac{2}{\pi}\dot{{r}}(t)\text{tan}^{-1}(\bar{\lambda}t)+\frac{2}{\pi}{r}(t)\frac{\bar{\lambda}}{1+\bar{\lambda}^2t^2} \notag \\
&-\text{sign}_\text{v}(s({x}^{(2)}(t),{x}^{(2)}(0),t))\rho,
\end{align}
where $ \rho\in R^r $ contains the upper bounds on the absolute values of the elements of the disturbance vector $ d $, and $ s({x}^{(2)}(t),{x}^{(2)}(0),t) $ is the sliding surface (\ref{eq:e52}). Then the trajectories of $ x^{(2)} $ stay on the sliding surface (\ref{eq:e52}).

\noindent
\textbf{Proof.} Consider a Lyapunov function candidate $V_2(s)=\frac{1}{2}s^Ts$, where the arguments of $ s({x}^{(2)}(t),{x}^{(2)}(t_0),t) $ are dropped for clarity. By taking the derivative of $V_2$, and using (\ref{eq:e52}) with $ t_0=0 $, we obtain,
\begin{align}\label{eq:e61xx}
\dot{V}_2=s^T\dot{s}&=s^T \Big(\dot{{x}}^{(2)}(t)+\bar{\lambda}{x}^{(2)}(0)e^{-\bar{\lambda}t}-\frac{2}{\pi}\dot{{r}}(t)\text{tan}^{-1}(\bar{\lambda}t)\notag \\
&-\frac{2}{\pi}{r}(t)\frac{\bar{\lambda}}{1+\bar{\lambda}^2t^2} \Big).
\end{align}
Using (\ref{eq:assump1}) and Assumption 2, we have $ \dot{x}^{(2)}(t)=A_{2,1}x^{(1)}(t)+A_{2,2}x^{(2)}(t)+v+d $. By substituting it  in (\ref{eq:e61xx}) we have 
\begin{align}\label{eq:e61x}
\dot{V}_2&=s^T\dot{s}=s^T \Big(A_{2,1}x^{(1)}(t)+A_{2,2}x^{(2)}(t)+v+d\notag \\
&+\bar{\lambda}{x}^{(2)}(0)e^{-\bar{\lambda}t}-\frac{2}{\pi}\dot{{r}}(t)\text{tan}^{-1}(\bar{\lambda}t)-\frac{2}{\pi}{r}(t)\frac{\bar{\lambda}}{1+\bar{\lambda}^2t^2} \Big).
\end{align}

By substituting the control law (\ref{eq:e60}) into (\ref{eq:e61x}), and using Definitions 1 and 2, it follows that
\begin{align}\label{eq:e62z}
\dot{V}_2&=
s^T[{d}-\text{sign}_\text{v}(s)\rho]=s^T{d}-|s^T|_\text{v}\rho\notag \\
&\leq |s^T|_\text{v}(|d|_\text{v}-\rho).
\end{align}
Since $ \rho $ contains the upper bounds on the absolute values of the elements of the disturbance vector $ d $, the elements of $ |d|_\text{v}-\rho $ are non-positive, which leads to $\dot{V}_2 \leq 0$, and consequently proves that $ x^{(2)} $ trajectories which are on sliding surface at $ t=t_0 $, will remain there for all $ t>0 $. 
\qed

\subsection{\textbf{{Bounding the control input}}}\label{saturation}

\begin{figure}
	\begin{center}
		\includegraphics[width=8.5cm]{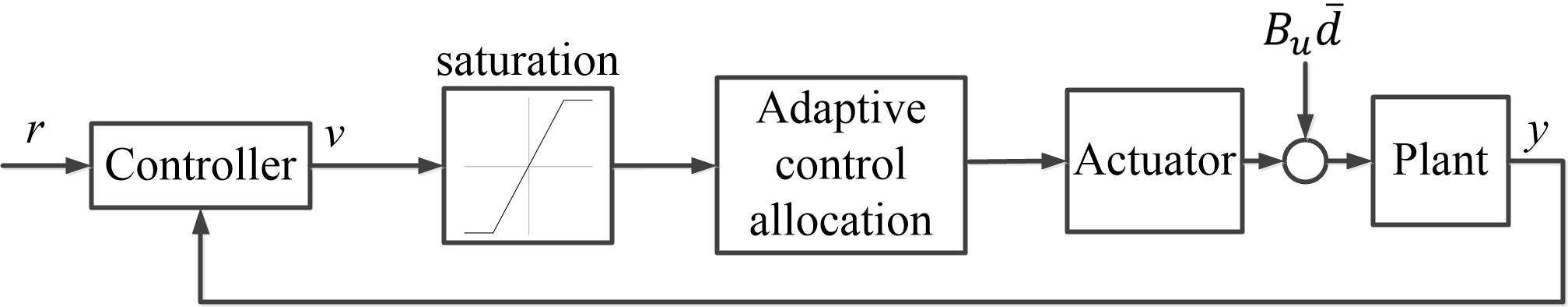}    %
		\caption{Block diagram of the closed loop system with soft saturation.} 
		\label{fig:Drawingxxx}
	\end{center}
\end{figure}

\noindent
\textbf{Remark 10.}
Up until now, we showed that the control signal $ v $ given in (\ref{eq:e60}) keeps $ x^{(2)} $ trajectories on the sliding surface defined in (\ref{eq:e52}), and as long as the states of the system remain on the sliding surface, the system output $ y $ follows the reference $ r $ while all the states remain bounded. In the control allocation development, we stated that the boundedness of the control signal $ v $ is ensured by using a soft saturation bound, the limits of which are set in Section \ref{project} as $ v_i\in [-M_i, M_i], i=1, ..., r $. The overall closed loop system block diagram is presented in Figure \ref{fig:Drawingxxx}. In this section, we provide a method inspired by \cite{CorCri10}, to make sure that the control signal $ v $ remains within this saturation bounds.

To ensure that $ |{v}_{i}|\leq {M}_i $ for $ i=1, ..., r $, the following inequality, obtained by using (\ref{eq:e60}), should be satisfied for all $ i=1, ..., r$,
\begin{align}\label{eq:e63xx}
|{v}_{i}(t)|&=\Big|-\sum_{j=1}^{n}a_{2_{i,j}}{x}_j(t)-\bar{\lambda}{x}_{n-r+i}(0)e^{-\bar{\lambda}t}\notag \\ &+\frac{2}{\pi}\dot{{r}}_i(t)\text{tan}^{-1}(\bar{\lambda }t)+\frac{2}{\pi}{r}_i(t)\frac{\bar{\lambda}}{1+\bar{\lambda}^2t^2}\notag \\
&-\bar{s}_i({x(t)})\rho_i\Big|\leq {M}_i,
\end{align}
where $ {x}_j $, $ {r}_i $, $ \dot{{r}}_i $ and $ \rho_i $ denote the $j$th component of $ x(t) $ and $ i $th components of $ {r}(t) $, $ \dot{{r}}(t) $ and $ \rho $, respectively. Moreover,  $\bar{{s}}_i({x(t)})$ is the $ i $th diagonal element of 
$\text{sign}_{\text{v}}(s({x}^{(2)}(t),{x}^{(2)}(0),t))$.  Defining $ A_1=[A_{1,1}\ A_{1,2}] $, $ A_2=[A_{2,1}\ A_{2,2}] $, where $ A_{i,j} $ is defined in (74), $ a_{2_{i,j}} $ refer to the elements of $ A_2 $. Using the inequalities $ |a-b|<|a|+|b| $ and $ |a+b|<|a|+|b| $ for two scalars $ a $ and $ b $, it can be shown that satisfying the following inequality, ensures (\ref{eq:e63xx}):
\begin{align}\label{eq:e64}
&\Big|\sum_{j=1}^{n}a_{2_{i,j}}{x}_j\Big|+\Big|\bar{\lambda}{x}_{n-r+i}(0)e^{-\bar{\lambda}t}\Big|+\Big|\frac{2}{\pi}\dot{{r}}_i\text{tan}^{-1}(\bar{\lambda }t)\Big|\notag \\
&+\Big|\frac{2}{\pi}{r}_i\frac{\bar{\lambda}}{1+\bar{\lambda}^2t^2}\Big|\leq ({M}_i-\rho_i).
\end{align}
Remembering that $ \rho_i $ is the upper bound of the disturbance element $ d_i $, and $ |d_i| $ is always smaller than $ M_i $ (see Remark 4), we obtain that $ (M_i-\rho_i)>0 $, meaning that the right hand side of the inequality (\ref{eq:e64}) is positive.
%

It can then be shown that satisfying the following inequality ensures (\ref{eq:e64})
\begin{align}\label{eq:e65}
\Big(\Big|\sum_{j=1}^{n}a_{2_{i,j}}{x}_j\Big|+\Big|\dot{{r}}_i\Big|-{M}_i+\rho_i\Big)&+\bar{\lambda}\Big(\Big|{x}_{n-r+i}(0)\Big|\notag \\
&+\frac{2}{\pi}\Big|{r}_i\Big|\Big)\leq 0.
\end{align}
Remembering that $ x^{(1)}=[x_1, ..., x_{n-r}]^T $ and $ x^{(2)}=[x_{n-r+1}, ..., x_{n}]^T $, and using (\ref{eq:e54}), it follows that $ |{x}_{n-r+i}(t)|\leq |{x}_{n-r+i}(0)|+|r_i(t)| $ for $ i=1, ..., r $, and $ ||x^{(1)}(t)||\geq x_j $ for all $ j=1, ...,n-r $. 
\begin{align}\label{eq:e65xx}
&\Big(\Big|\sum_{j=1}^{n-r}a_{2_{i,j}}{x}_j\Big|+\Big|\sum_{j=n-r+1}^{n}a_{2_{i,j}}\big( |{x}_{n-r+i}(0)|+|r_i| \big)\Big|\notag \\
&+\Big|\dot{{r}}_i\Big|-{M}_i+\rho_i\Big)+\bar{\lambda}\Big(\Big|{x}_{n-r+i}(0)\Big|+\frac{2}{\pi}\Big|{r}_i\Big|\Big)\leq 0.
\end{align}
Furthermore, the upper bound of $ ||x^{(1)}(t)|| $ is obtained in Lemma 4. Therefore, defining $ \bar{x}^{(1)}(0) $, $ \bar{x}^{(2)}(0) $, $ \bar{{r}} $, $ \bar{r}_i $ and $ \bar{\dot{r}}_i $ as bounds on $ ||x^{(1)}(0)|| $, $ ||x^{(2)}(0)|| $, $ ||r(t)|| $, $ |r_i(t)|$ and $ |\dot{r}_i(t)| $, it can be shown that satisfying the following inequality ensures (\ref{eq:e65}):
\begin{align}\label{eq:e65x}
&\Big(\Big|\sum_{j=1}^{n-r}a_{2_{i,j}}\Big|\big(k\bar{x}^{(1)}(0)+(1+K_2)\bar{x}^{(2)}(0)+K_2\bar{{r}}+\bar{{r}}_i\big)\notag \\
&+\bar{\dot{r}}_i-{M}_i+\rho_i\Big)+\bar{\lambda}\Big(\Big|\bar{x}^{(2)}(0)\Big|+\frac{2}{\pi}\Big|\bar{r}_i\Big|\Big)\leq 0,
\end{align}
where $ k $ and $ K_2 $ are defined in Lemma 4.
Equation (\ref{eq:e65x}) can be rewritten as
\begin{align}\label{eq:e66}
W_{i,1}+\bar{\lambda}W_{i,2}\leq 0,
\end{align}
where $  W_{i,1} $ is the first term, and $  W_{i,2} $ is the term multiplying $ \bar{\lambda } $ in (\ref{eq:e65x}). Note that $ W_{i,1} $ and $ W_{i,2} $ are functions of $ \bar{x}^{(1)}(0) $, $ \bar{x}^{(2)}(0) $, $ \bar{{r}}_i $, and $ \bar{\dot{r}}_i $ and remain constant along the closed-loop trajectory. Since $ W_{i,2} $ is positive, a value of $ \bar{\lambda}>0 $ satisfying (\ref{eq:e66}) can always be found if $ W_{i,1}<0 $, which can be realized by putting suitable bounds on $ \bar{x}^{(1)}(0) $, $ \bar{x}^{(2)}(0) $, $ \bar{r} $, $ \bar{r}_i $ and $ \bar{\dot{{r}}}_i $. A step by step design procedure to determine the controller parameters is given in the Appendix.

\section{SIMULATION RESULTS}


The Aerodata Model in Research Environment (ADMIRE), which represents the dynamics of an over-actuated aircraft model, is used to demonstrate the effectiveness of the adaptive control allocation in the presence of uncertainty. The linearized ADMIRE model is introduced in \cite{Har05}, and is given below:
\begin{equation}\label{eq:e59}
\begin{array}{ll}
x=[\alpha \ \beta \ p \ q \ r]^T\\
y=[p \ q \ r]^T\\
u=[u_c \ u_{re} \ u_{le} \ u_r]^T\\
\dot{x}=Ax+B_u u=Ax+B_vv \\
v=Bu,\ \ \  B_u=B_vB,\ \ \  B_v=\left[ \begin{array}{ll} 0_{2\times3} \\ I_{3\times3} \end{array}\right],
\end{array}
\end{equation}
where $\alpha, \beta, p, q$ and $r$ are the angle of attack, sideslip angle, roll rate, pitch rate and yaw rate, respectively. $u$ represents the control surface deflections vector which consists of canard wings, right and left elevons and the rudder. The state and control matrices are given by:
%
%
%
\begin{equation}\label{eq:e61}
\begin{array}{ll}
A=\\ \left[ \begin{array}{cccccc}-0.5432 & 0.0137 & 0 & 0.9778 & 0 \\ 0 & -0.1179 & 0.2215 & 0 & -0.9661 \\  0 & -10.5123 & - 0.9967 & 0 & 0.6176 \\  2.6221 & -0.0030 & 0 & -0.5057 & 0 \\  0 & 0.7075 & -0.0939 & 0 & -0.2127\end{array} \right],
\end{array}
\end{equation}
\begin{equation}\label{eq:e62}
\begin{array}{ll}
B_u=\left[ \begin{array}{cccccccccccc}0.0069 & -0.0866 & -0.0866 & 0.0004 \\ 0 & 0.0119 & -0.0119 & 0.0287 \\0 & -4.2423 & 4.2423 & 1.4871 \\ 1.6532 & -1.2735 & -1.2735 & 0.0024 \\  0 & -0.2805 & 0.2805 & -0.8823\end{array} \right].
\end{array}
\end{equation}
The position limits of the control surfaces are given as
\begin{equation}\label{eq:e63}
\begin{array}{ll}
u_c\in[-55,25]\times \frac{\pi}{180}rad,\ \ u_{re},u_{le},u_r\in[-30,30]\times \frac{\pi}{180}rad, \notag
\end{array}
\end{equation}
%
with first-order dynamics and a time constant of $ 0.05(sec) $. Note that the control surfaces influence on derivatives of the first two states i.e. $\dot{\alpha}$ and $\dot{\beta}$  is neglected, that is the control surfaces are considered to be pure moment generators, so that control allocation implementation becomes possible \cite{Har05}. 

To represent actuator loss of effectiveness and disturbance, a diagonal matrix $\Lambda$ and a vector $ d_u $, respectively, are augmented to the model (\ref{eq:e59}) as
\begin{equation}\label{eq:e59xxx}
\begin{array}{ll}
\dot{x}=Ax+B_u\Lambda u+ B_ud_u=Ax+B_vv+B_v\bar{d} \\
v=B\Lambda u,\ \ \  \bar{d}=Bd_u, \ \ \ B_u=B_vB,\ \ \  B_v=\left[ \begin{array}{ll} 0_{2\times3} \\ I_{3\times3} \end{array}\right].
\end{array}
\end{equation}

The system (\ref{eq:e59xxx}) with state and input matrices (\ref{eq:e62}) can be decomposed into two subsystems:
\begin{equation}\label{eq:e62x}
\begin{array}{ll}
\left[ \begin{array}{l} \dot{\alpha}\\ \dot{\beta}\end{array}\right]&=\left[ \begin{array}{cccc} -0.5432 & 0.0137\\  0 & -0.1179 \end{array}\right]\left[ \begin{array}{l} \alpha\\\beta \end{array}\right]\\
&+\left[ \begin{array}{cccc} 0 & 0.9778 & 0\\ 0.2215 & 0 & -0.9661 \end{array}\right]\left[ \begin{array}{l} p\\q \\ r \end{array}\right],\notag\\
\left[ \begin{array}{l} \dot{p}\\ \dot{q}\\ \dot{r}\end{array}\right]&=\left[ \begin{array}{cccc} 0 & -10.5123\\  2.6221 & -0.0030 \\ 0 & 0.7075 \end{array}\right]\left[ \begin{array}{l} \alpha\\\beta \end{array}\right] \\
&+\left[ \begin{array}{cccc} - 0.9967 & 0 & 0.6176\\ 0 & -0.5057 & 0\\ -0.0939 & 0 & -0.2127 \end{array}\right]\left[ \begin{array}{l} p\\q \\ r \end{array}\right]+v+\bar{d}. \notag
\end{array}
\end{equation}
These two subsystem representation satisfy the assumptions required for implementing the sliding mode controller design in the previous section. In the simulations, the disturbance $ \bar{d} $ is a sinusoidal function with amplitude $ 0.1 $ and frequency $ 1 $ rad/s while a zero-mean white Gaussian noise with standard deviation $ \sigma_x=0.0035 $ rad represents the measurement noise. Following the steps in the Appendix, the controller design parameter $ \bar{\lambda} $ is calculated as $ \bar{\lambda}=3 $. In addition, to avoid chattering, a boundary layer approach is implemented \cite{Che02}.

\begin{figure}
	\begin{center}
		\includegraphics[width=9.0cm]{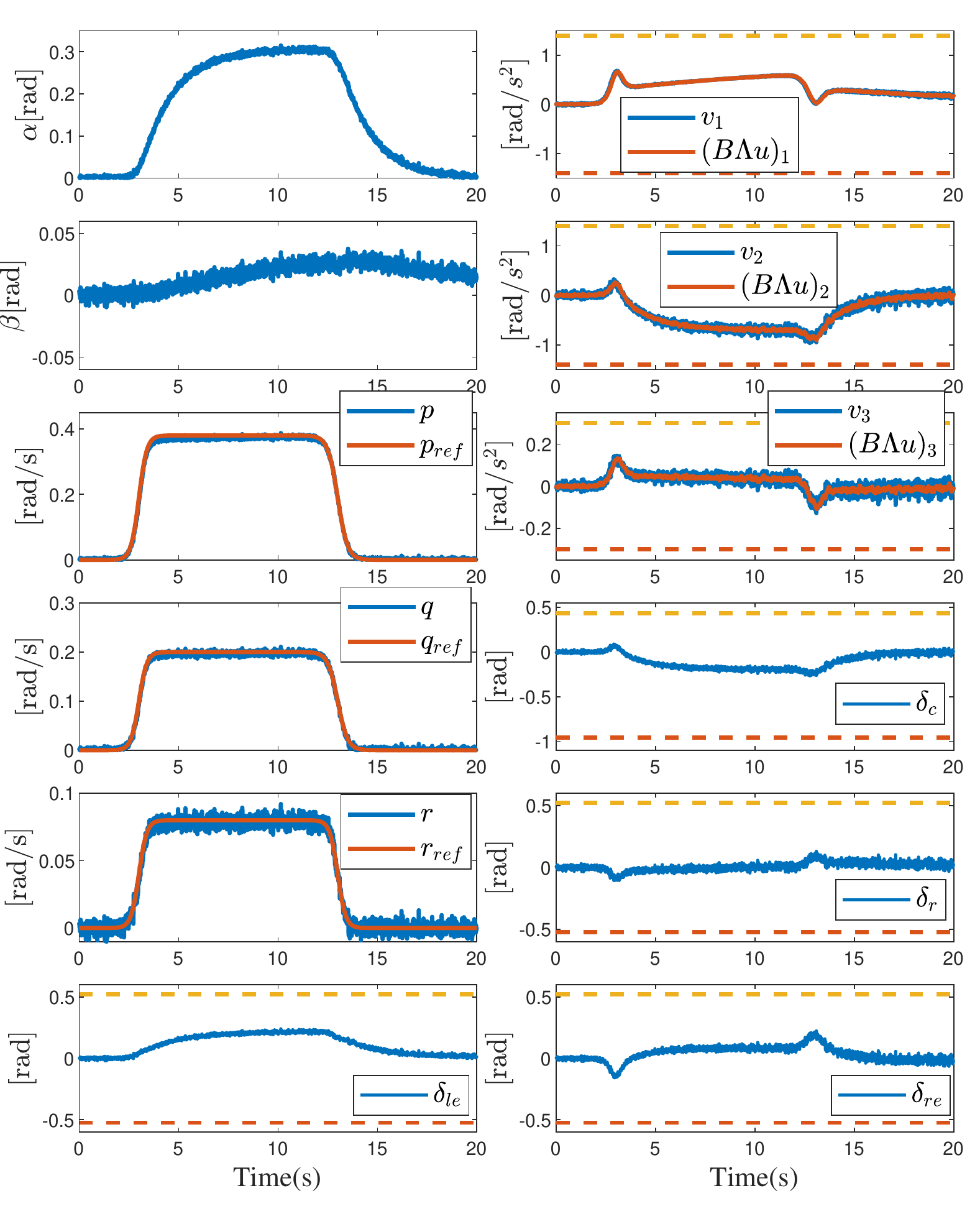}    %
		\caption{System states, virtual control signals and actuators' deflections using conventional control allocation when actuator loss of effectiveness is  $ \Lambda_1=I $.} 
		\label{fig:simulpseudonofault}
	\end{center}
\end{figure}

The simulation results for a conventional pseudo inverse based control allocation are reported in Figure \ref{fig:simulpseudonofault} for the case without actuator loss of effectiveness, that is,  $ \Lambda_1=I $.

\begin{figure}
	\begin{center}
		\includegraphics[width=9.0cm]{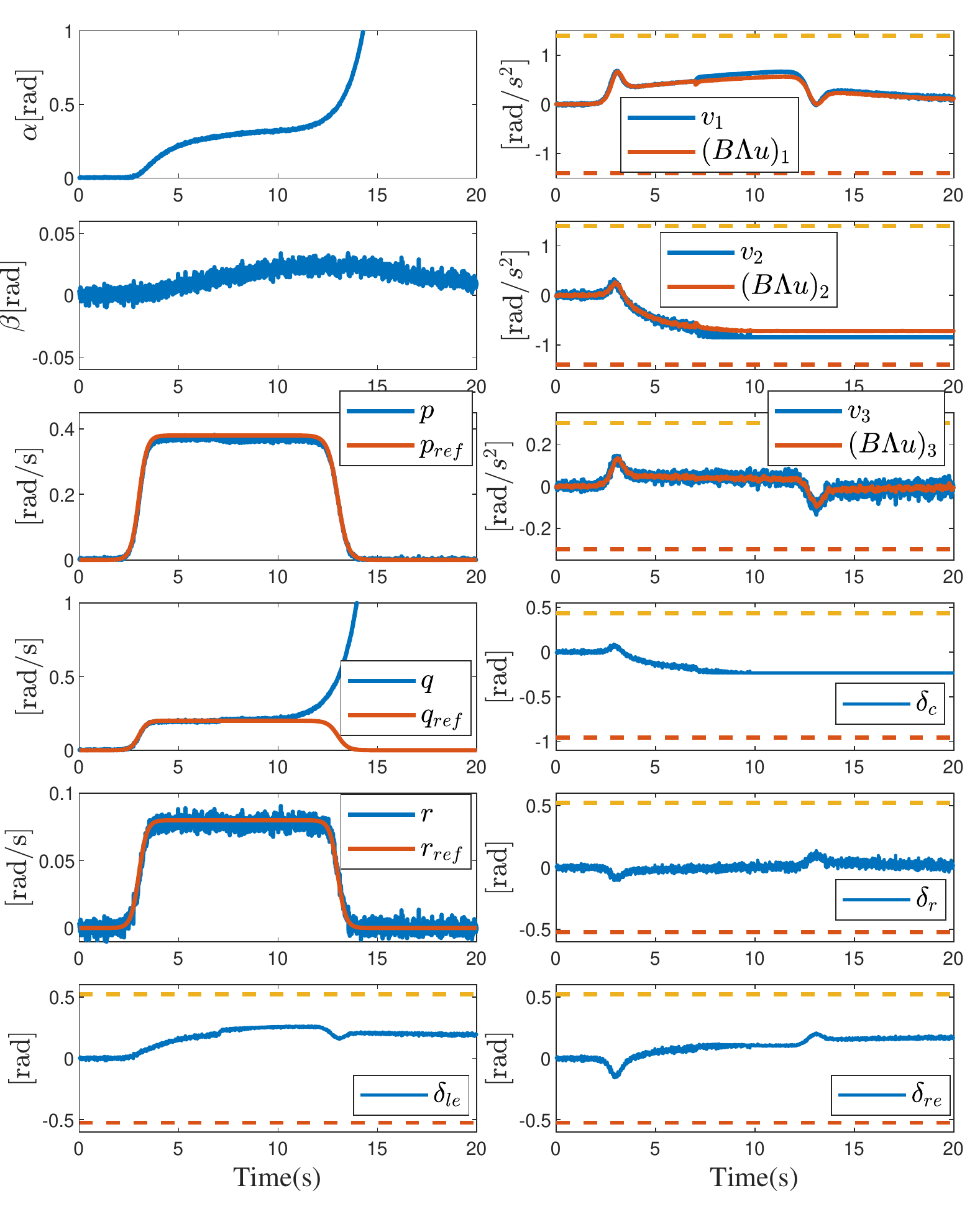}    %
		\caption{System states, virtual control signals and actuators' deflections using conventional control allocation when actuator loss of effectiveness is  $ \Lambda_2 $.} 
		\label{fig:simulpseudofault}
	\end{center}
\end{figure}

With the actuator loss of effectiveness modeled as
\begin{equation}\label{eq:e62xxxx}
\Lambda_2(t)=\left \{\begin{array}{ll}diag(1, 1, 1, 1)&for\  t<7(sec),\\ diag(0.85, 0.85, 0.85, 0.85)&for\  t\geq 7(sec),\end{array}\right.\notag
\end{equation}
the simulation results for the conventional control allocation are given in Figure \ref{fig:simulpseudofault}. It is seen that $ 15\% $ loss of effectiveness in all actuators at $ t=7 $ sec causes instability.

The proposed adaptive control allocation introduced in Section III will be used in the following simulations. Using Steps 1-3 in Section IV, the values of $ M_1=1.4 $, $ M_2=1.4 $ and $ M_3=0.3
$ are obtained and $ \Omega_{\text{proj}} $, which determines the maximum and minimum of each element of the adaptive parameter matrix is computed corresponding to $ \theta_{v_{1,1}}\in [-0.0129, 0.0129] $, $ \theta_{v_{1,2}}\in [0.0307,0.5225] $, $ \theta_{v_{1,3}}\in [-0.1357, 0.1371] $, $ \theta_{v_{1,4}}\in [-0.212, 0] $, $ \theta_{v_{2,1}}\in [-0.3149, -0.1113] $, $ \theta_{v_{2,2}}\in [-0.217, -0.1416] $, $ \theta_{v_{2,3}}\in [-0.0241, 0.2363] $, $ \theta_{v_{2,4}}\in [-0.4162, -0.01] $, $ \theta_{v_{3,1}}\in [0.1587, 0.1977] $,
$ \theta_{v_{3,2}}\in [0.0673, 0.0675] $,
$ \theta_{v_{3,3}}\in [-0.001, 0.001] $,
$ \theta_{v_{3,4}}\in [-1.2755, -0.7641] $.
We use a closed loop reference model with $l=4$ and $A_m$ selected as
\begin{equation}\label{eq:e62xxx}
A_m=-\left[ \begin{array}{cccccccccccc}0.2 & 0 & 0 \\ 0 & 0.1 & 0 \\  0 & 0 & 0.1\end{array} \right].\notag
\end{equation}

\begin{figure}
	\begin{center}
		\includegraphics[width=9.0cm]{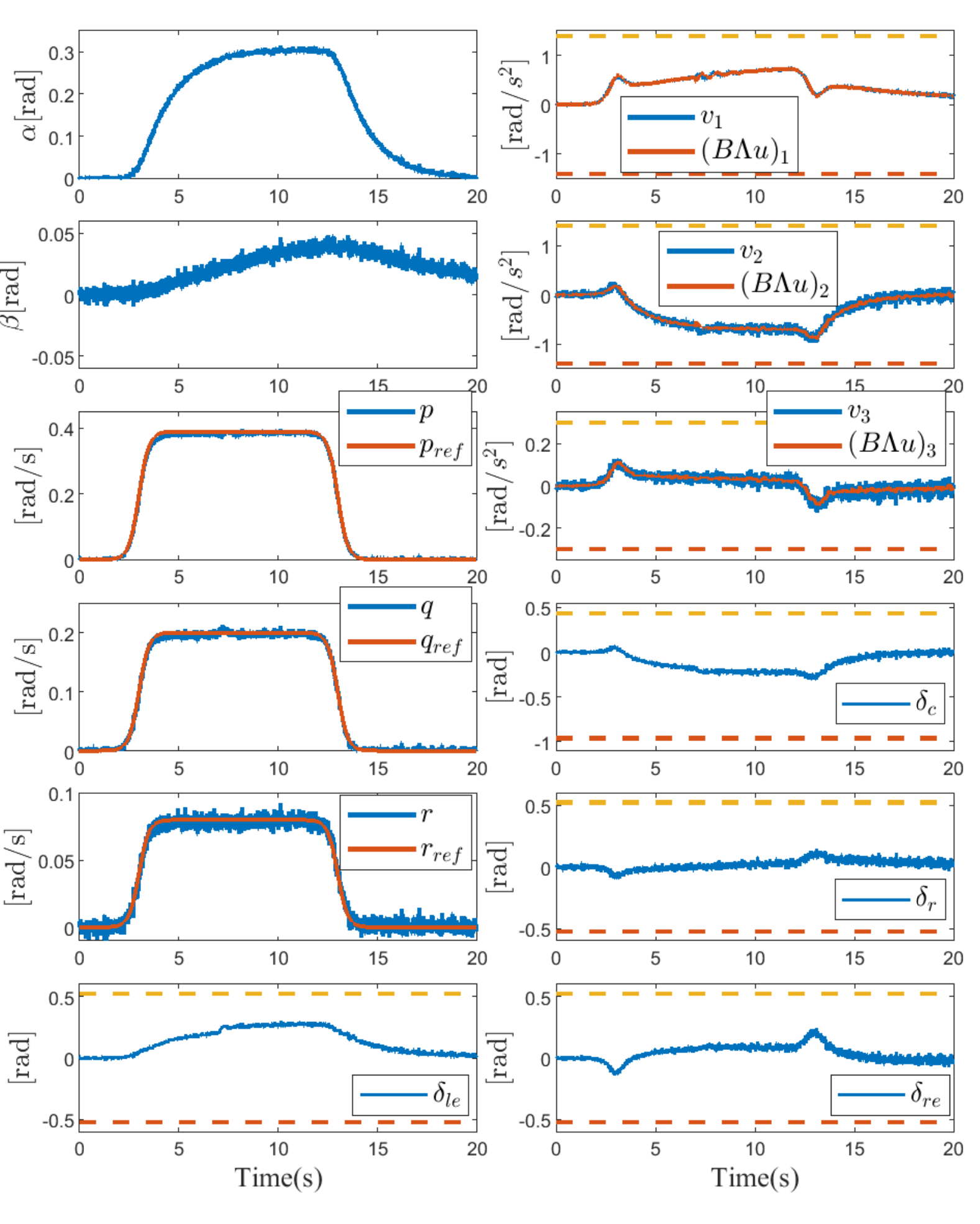}    %
		\caption{System states, virtual control signals and actuators' deflections using adaptive control allocation, when actuator loss of effectiveness matrix is $ \Lambda_2 $.} 
		\label{fig:simulACAfault1}
	\end{center}
\end{figure}

Figure \ref{fig:simulACAfault1} shows the simulation results for the system with our adaptive control allocation and the actuator loss of effectiveness matrix $ \Lambda_2(t) $.

\begin{figure}
	\begin{center}
		\includegraphics[width=9.0cm]{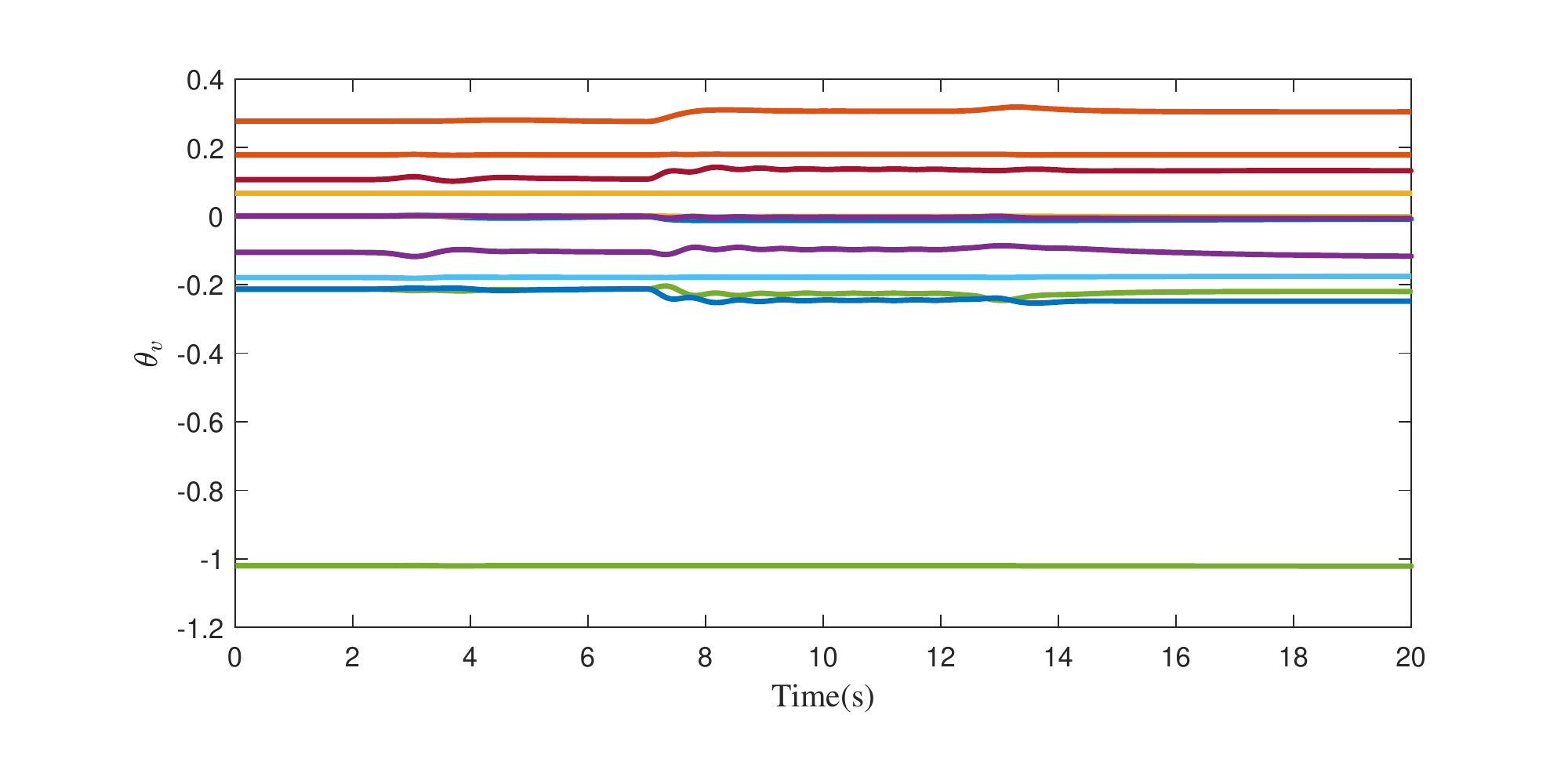}    %
		\caption{Adaptive parameters when the actuator loss of effectiveness matrix is $ \Lambda_2 $.} 
		\label{fig:teta1}
	\end{center}
\end{figure}

\begin{figure}
	\begin{center}
		\includegraphics[width=9.0cm]{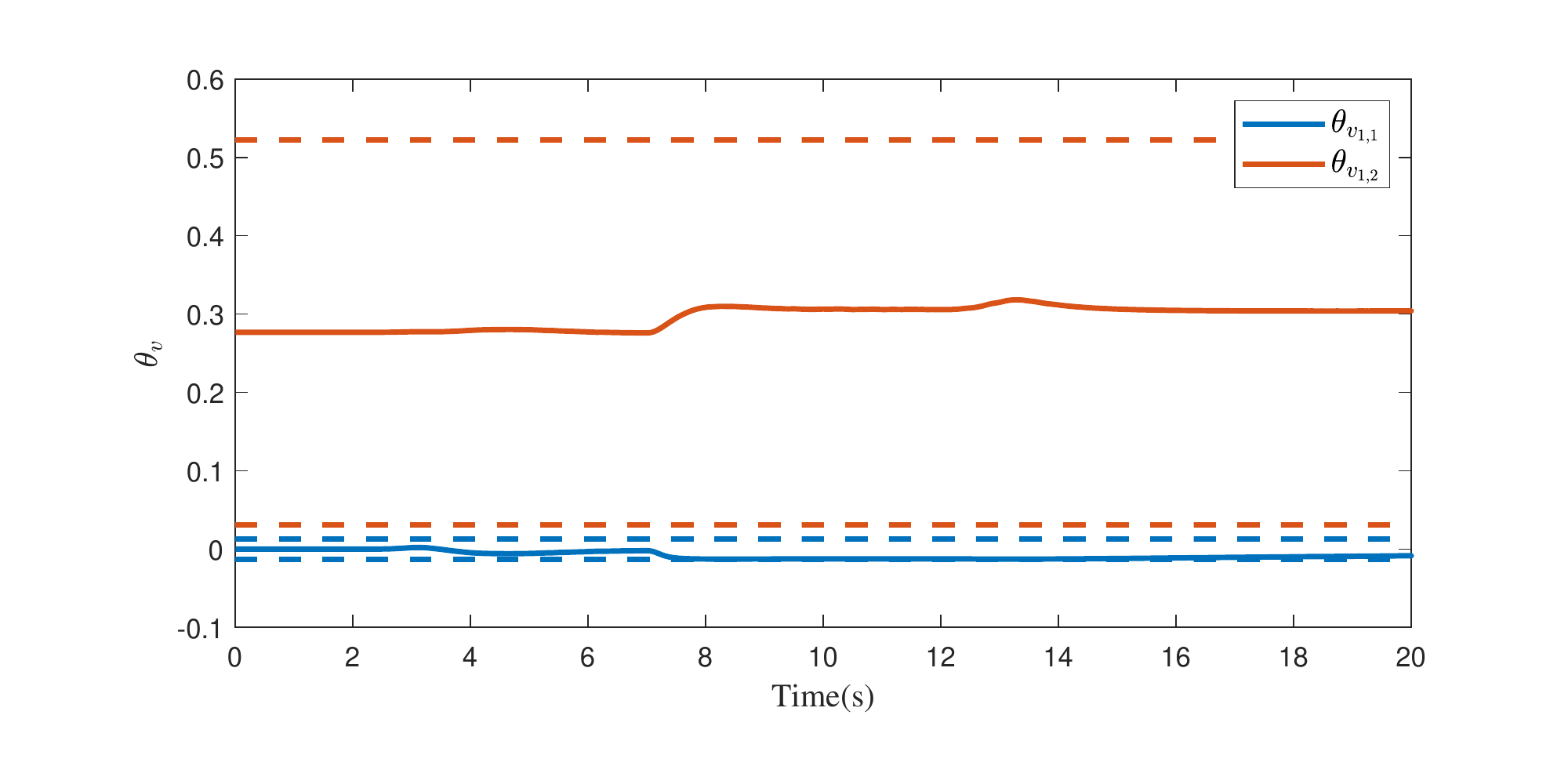}    %
		\caption{Two adaptive parameters and their projection boundaries when the actuator loss of effectiveness matrix is $ \Lambda_2 $. The dashed lines are the projection boundaries.} 
		\label{fig:tetproj1}
	\end{center}
\end{figure}

It is seen that the first two states, $ \alpha $ and $ \beta $, are bounded, and the other states ($ p $, $ q $ and $ r $) follow the reference inputs ($ p_{ref} $, $ q_{ref} $ and $ r_{ref} $) even after the introduction of $ 30\% $ actuator loss of effectiveness at $ t=7 $ sec. Also, it is seen that the elements of $ (B\Lambda u)_i $ for $ i=1, 2, 3 $, converge to the virtual control signal elements $ v_i $ for $ i=1, 2, 3 $. The time histories of the adaptive parameters, which are the elements of $ \theta_v $ matrix, are shown in Figure \ref{fig:teta1}. Two adaptive parameters are selected to illustrate their deviation inside their projection boundaries in Figure \ref{fig:tetproj1}. 

\begin{figure}
	\begin{center}
		\includegraphics[width=9.0cm]{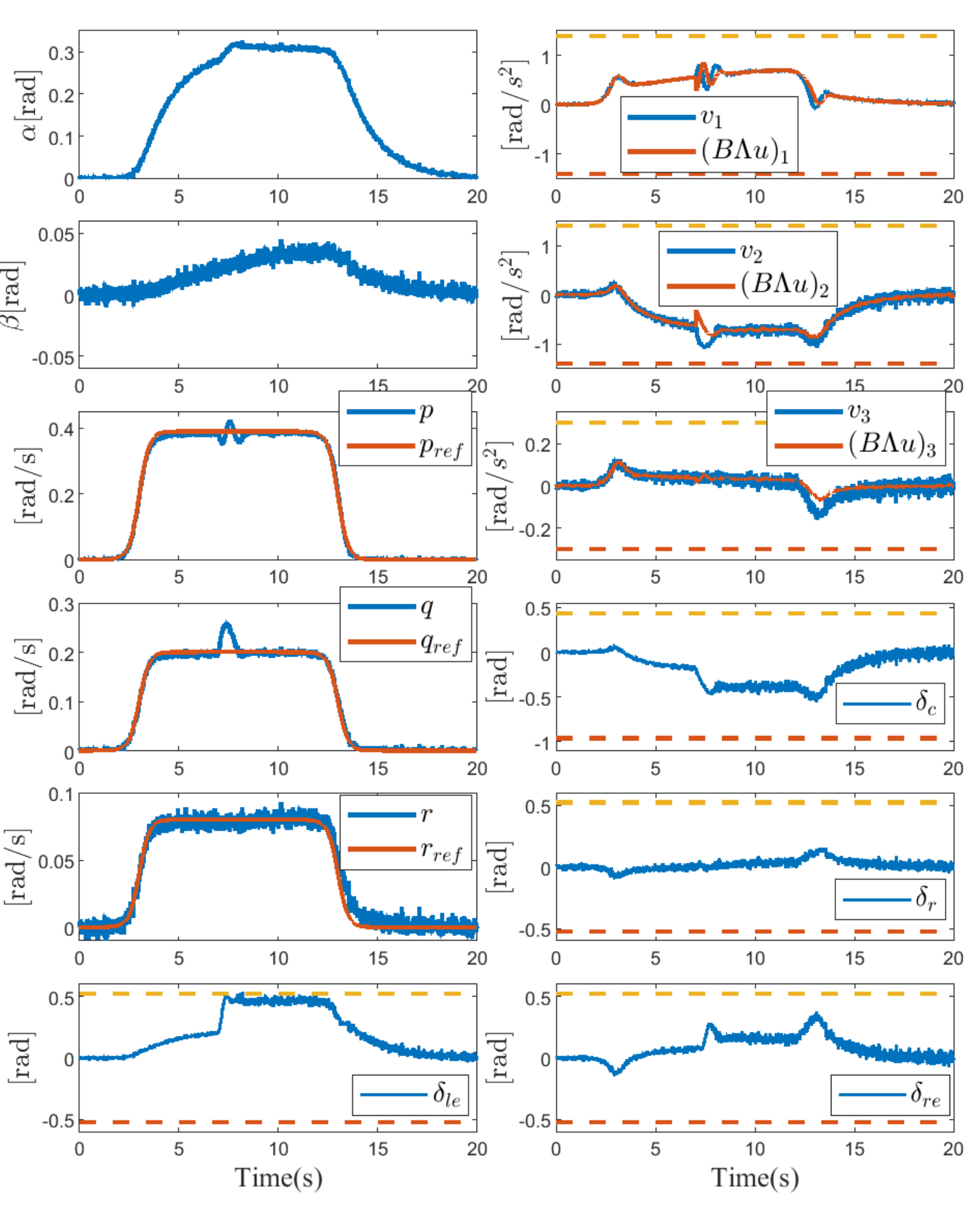}    %
		\caption{System states, virtual control signals and actuators' deflections using adaptive control allocation when actuator loss of effectiveness matrix is $ \Lambda_3 $.} 
		\label{fig:simulACAfault2}
	\end{center}
\end{figure}

\begin{figure}
	\begin{center}
		\includegraphics[width=9.0cm]{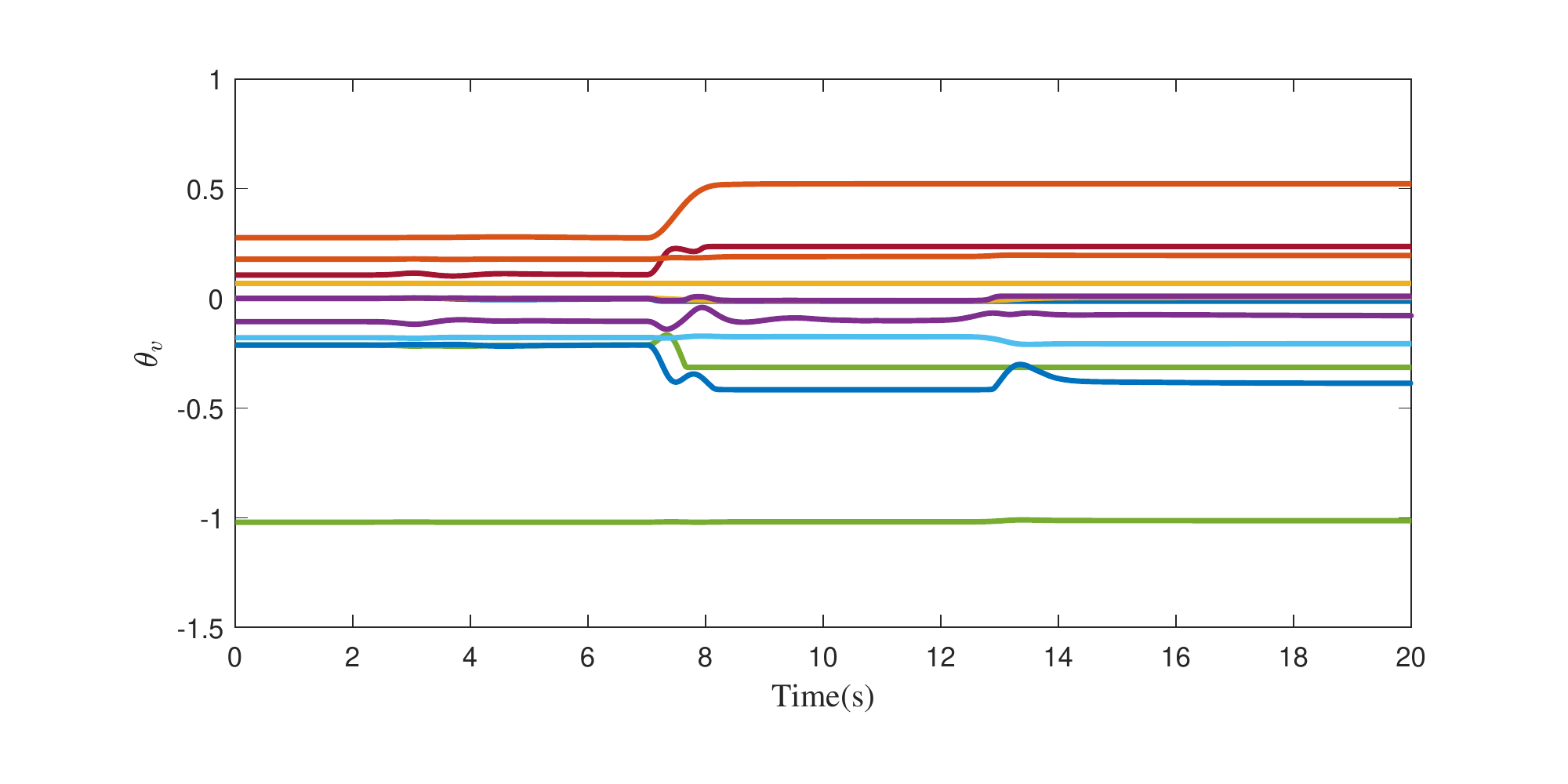}    %
		\caption{Adaptive parameters when the actuator loss of effectiveness matrix is $ \Lambda_3 $.} 
		\label{fig:teta2}
	\end{center}
\end{figure}

\begin{figure}
	\begin{center}
		\includegraphics[width=9.0cm]{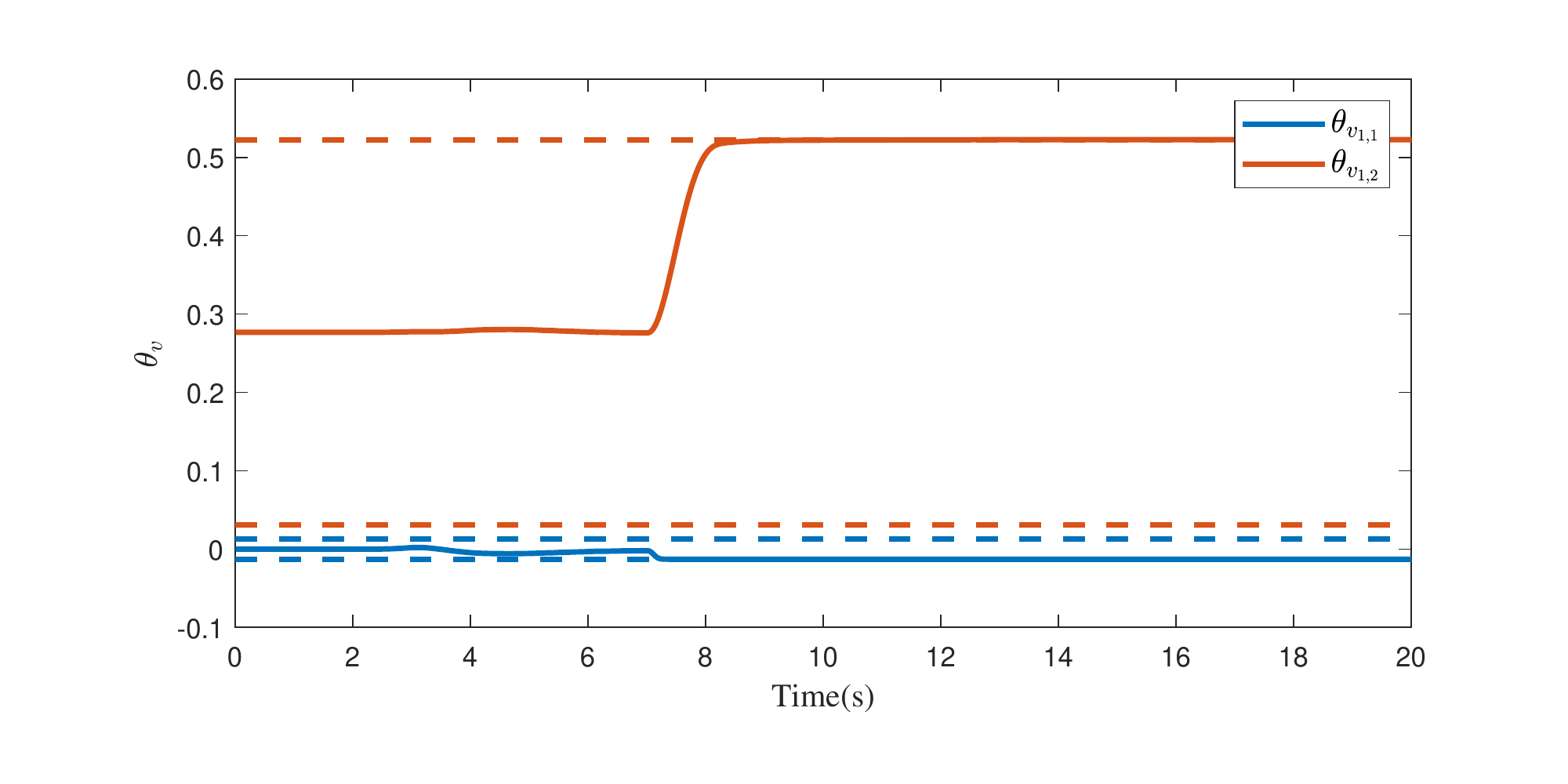}    %
		\caption{Two adaptive parameters and their projection boundaries when the actuator loss of effectiveness matrix is $ \Lambda_3 $. The dashed lines are the projection boundaries.} 
		\label{fig:tetproj2}
	\end{center}
\end{figure}

Another scenario is considered next, where a $ 50\% $ loss of effectiveness for the control surfaces are simulated, specifically, with
\begin{equation}\label{eq:e62xxxxx}
\Lambda_3(t)=\left \{\begin{array}{ll}diag(1, 1, 1, 1)&for\  t<7(sec),\\ diag(0.5, 0.5, 0.5, 0.5)&for\  t\geq 7(sec).\end{array}\right.\notag
\end{equation}
It is seen in Figure \ref{fig:simulACAfault2} that the system remains stable after the introduction of the loss of effectiveness. The time histories of the adaptive parameters, which are the elements of $ \theta_v $ matrix, are shown in Figure \ref{fig:teta2}. Moreover, two adaptive parameters are selected to illustrate their deviation within their projection boundaries in Figure \ref{fig:tetproj2}. 

%





\section{SUMMARY}
An adaptive control allocation for uncertain over-actuated systems with actuator saturation is proposed in this paper. The method needs neither uncertainty identification nor persistence of excitation. A sliding mode controller with time-varying sliding surface is also proposed, to guarantee the stability of the overall closed loop system while realizing reference tracking. The simulation results with the ADMIRE model show the effectiveness of the proposed method.  


\appendix

\section{Controller design procedure}
The following procedure can be followed to obtain the controller design parameters: 

\noindent
\textbf{Step 1-} Use Step 1 in Section IV, to determine $ M_i $, $ i=1, ..., r $. 

\noindent
\textbf{Step 2-} Calculate $ \Omega_{\theta} $ using Step 2 in Section IV.


%


\noindent
\textbf{Step 3-} Using Theorem 4, calculate $ \gamma\equiv\max_{i}(1-\sqrt{\frac{\gamma_{M_i}}{\gamma_{B_i}}}) $, where $ M_{\text{max}}=\max_{i}{M_i} $, $ \gamma_{M_i}\equiv \frac{M_i^2}{M_{\text{max}}^2}-\epsilon$ for a small positive $ \epsilon $, and $ \gamma_{B_i}\equiv ||\text{row}_i(B)||||B^T(BB^T)^{-1}||$. 

\noindent
\textbf{Step 4-} Using $ \gamma $, obtain $ \Omega_{\Lambda_1} $ in (\ref{eq:lamda}).

\noindent
\textbf{Step 5-} Solve the optimization problem (\ref{eq:e41_1xxxx}), which leads to obtaining the $ \Omega_{\text{proj}} $, which is defined in Section IV, step 3.

\noindent
\textbf{Step 6-} Using (\ref{eq:lamda}) and $ \Omega_{\text{proj}} $, calculate $ \bar{\rho}_i $ for $ i=1, ..., r $, which are defined in Remark 4.

\noindent
\textbf{Step 9-} Calculate $ k $ and $ \xi $ using (\ref{eq:e57}).

\noindent
\textbf{Step 10-} In the proposed controller design, since the controllers' goal is reference tracking in the presence of saturation, the initial states, $ {x}^{(1)}(0) $ and $ {x}^{(2)}(0) $, the elements of reference input, $ r $, the elements of the derivative of the reference input, $ \dot{r} $, and the elements of disturbance $ \bar{d} $, should be bounded. Let $ \bar{x}^{(1)}(0) $, $ \bar{x}^{(2)}(0) $, $ \bar{r}_i $, $ \bar{\dot{r}}_i $ and $ \bar{L}_i $ be upper bounds on the norms of $ {x}^{(1)}(0) $, $ {x}^{(2)}(0) $, $ {r}_i $, $ {\dot{r}}_i $ and $ \bar{d}_i $, respectively, for $ i=1, ..., r $. These values can be obtained by an expert, who has information about the plant and its constraints. Also, these values can be written in the data sheets.

\noindent
\textbf{Step 11-} Check if $ W_{i,1}<0 $.

\textbf{If yes}, continue to the next step. 

\textbf{If no}, the expert should reduce $ \bar{x}^{(1)}(0) $, $ \bar{x}^{(2)}(0) $, $ \bar{r}_i $, $ \bar{\dot{r}}_i $ and $ \bar{L}_i $ to satisfy this inequality. Most of the time, this is done by confining the initial value of states to a smaller limit. In this design, it can be done even by reducing the derivative of the reference input, if fast maneuver is not required in the plant.

%
%
%
%
%
%

\noindent
\textbf{Step 12-} Calculate $ {W}_{i,2} $ in (\ref{eq:e66}).
%
%
%
%
%
%
%

\noindent
\textbf{Step 13-} Find a $ \bar{\lambda} $ satisfying (\ref{eq:e66}) and design the control signal (\ref{eq:e60}).

\bibliographystyle{plain}        
\bibliography{References}           

\begin{thebibliography}{10}

\bibitem{AcoYil14}
Diana~M Acosta, Yildiray Yildiz, Robert~W Craun, Steven~D Beard, Michael~W
  Leonard, Gordon~H Hardy, and Michael Weinstein.
\newblock Piloted evaluation of a control allocation technique to recover from
  pilot-induced oscillations.
\newblock {\em Journal of Aircraft}, 52(1):130--140, 2014.

\bibitem{Alw08}
Halim Alwi and Christopher Edwards.
\newblock Fault tolerant control using sliding modes with on-line control
  allocation.
\newblock {\em Automatica}, 44(7):1859--1866, 2008.

\bibitem{Ant06}
Panos~J Antsaklis and Anthony~N Michel.
\newblock {\em Linear systems}.
\newblock Springer Science \& Business Media, 2006.

\bibitem{Bod02}
Marc Bodson.
\newblock Evaluation of optimization methods for control allocation.
\newblock {\em Journal of Guidance, Control, and Dynamics}, 25(4):703--711,
  2002.

\bibitem{Bor96}
Kenneth~A Bordignon.
\newblock {\em Constrained control allocation for systems with redundant
  control effectors}.
\newblock PhD thesis, Virginia Tech, 1996.

\bibitem{Bou17}
Abdelkader Bouarfa, Marc Bodson, and Maurice Fadel.
\newblock A fast active-balancing method for the 3-phase multilevel flying
  capacitor inverter derived from control allocation theory.
\newblock {\em IFAC-PapersOnLine}, 50(1):2113--2118, 2017.

\bibitem{Buf96}
James~M Buffington and Dale~F Enns.
\newblock Lyapunov stability analysis of daisy chain control allocation.
\newblock {\em Journal of Guidance, Control, and Dynamics}, 19(6):1226--1230,
  1996.

\bibitem{CasGar10}
Alessandro Casavola and Emanuele Garone.
\newblock Fault-tolerant adaptive control allocation schemes for overactuated
  systems.
\newblock {\em International Journal of Robust and Nonlinear Control},
  20(17):1958--1980, 2010.

\bibitem{Che02}
Min-Shin Chen, Yean-Ren Hwang, and Masayoshi Tomizuka.
\newblock A state-dependent boundary layer design for sliding mode control.
\newblock {\em IEEE transactions on automatic control}, 47(10):1677--1681,
  2002.

\bibitem{Che13}
Mou Chen, Shuzhi~Sam Ge, Bernard Voon~Ee How, and Yoo~Sang Choo.
\newblock Robust adaptive position mooring control for marine vessels.
\newblock {\em IEEE Transactions on Control Systems Technology},
  21(2):395--409, 2013.

\bibitem{CorCri16}
Maria~Letizia Corradini and Andrea Cristofaro.
\newblock A nonlinear fault-tolerant thruster allocation architecture for
  underwater remotely operated vehicles.
\newblock {\em IFAC-PapersOnLine}, 49(23):285--290, 2016.

\bibitem{CorCri10}
Maria~Letizia Corradini, Andrea Cristofaro, and Giuseppe Orlando.
\newblock Robust stabilization of multi input plants with saturating actuators.
\newblock {\em IEEE Transactions on Automatic Control}, 55(2):419--425, 2010.

\bibitem{CriJoh14}
Andrea Cristofaro and Tor~Arne Johansen.
\newblock Fault tolerant control allocation using unknown input observers.
\newblock {\em Automatica}, 50(7):1891--1897, 2014.

\bibitem{CriJoh15}
Andrea Cristofaro, Marios~M Polycarpou, and Tor~Arne Johansen.
\newblock Fault diagnosis and fault-tolerant control allocation for a class of
  nonlinear systems with redundant inputs.
\newblock In {\em 2015 54th IEEE Conference on Decision and Control (CDC)},
  pages 5117--5123. IEEE, 2015.

\bibitem{Dem13}
Murat Demirci and Metin Gokasan.
\newblock Adaptive optimal control allocation using lagrangian neural networks
  for stability control of a 4ws--4wd electric vehicle.
\newblock {\em Transactions of the Institute of Measurement and Control},
  35(8):1139--1151, 2013.

\bibitem{Duc09}
Guillaume~JJ Ducard.
\newblock {\em Fault-tolerant flight control and guidance systems: Practical
  methods for small unmanned aerial vehicles}.
\newblock Springer Science \& Business Media, 2009.

\bibitem{Dur17}
Wayne Durham, Kenneth~A. Bordignon, and Roger Beck.
\newblock {\em Aircraft Control allocation}.
\newblock Springer Science \& Business Media, 2009.

\bibitem{Dur93}
Wayne~C Durham.
\newblock Constrained control allocation.
\newblock {\em Journal of Guidance, control, and Dynamics}, 16(4):717--725,
  1993.

\bibitem{Edw10}
Christopher Edwards, Thomas Lombaerts, Hafid Smaili, et~al.
\newblock Fault tolerant flight control.
\newblock {\em Lecture Notes in Control and Information Sciences}, 399:1--560,
  2010.

\bibitem{FalHol16}
Guillermo~P Falcon{\'\i} and Florian Holzapfel.
\newblock Adaptive fault tolerant control allocation for a hexacopter system.
\newblock In {\em American Control Conference (ACC), 2016}, pages 6760--6766.
  IEEE, 2016.

\bibitem{Gal18}
Sergio Galeani and Mario Sassano.
\newblock Data-driven dynamic control allocation for uncertain redundant
  plants.
\newblock In {\em 2018 IEEE Conference on Decision and Control (CDC)}, pages
  5494--5499. IEEE, 2018.

\bibitem{GibAnnLav15}
Travis~E Gibson, Zheng Qu, Anuradha~M Annaswamy, and Eugene Lavretsky.
\newblock Adaptive output feedback based on closed-loop reference models.
\newblock {\em IEEE Transactions on Automatic Control}, 60(10):2728--2733,
  2015.

\bibitem{Gie06}
Witold Gierusz and Miroslaw Tomera.
\newblock Logic thrust allocation applied to multivariable control of the
  training ship.
\newblock {\em Control Engineering Practice}, 14(5):511--524, 2006.

\bibitem{Har02}
Ola H{\"a}rkeg{\aa}rd.
\newblock Efficient active set algorithms for solving constrained least squares
  problems in aircraft control allocation.
\newblock In {\em Decision and Control, 2002, Proceedings of the 41st IEEE
  Conference on}, volume~2, pages 1295--1300. IEEE, 2002.

\bibitem{Har05}
Ola H{\"a}rkeg{\aa}rd and S~Torkel Glad.
\newblock Resolving actuator redundancy—optimal control vs. control
  allocation.
\newblock {\em Automatica}, 41(1):137--144, 2005.

\bibitem{JohFos13}
Tor~A Johansen and Thor~I Fossen.
\newblock Control allocation—a survey.
\newblock {\em Automatica}, 49(5):1087--1103, 2013.

\bibitem{Joh08}
Tor~A Johansen, Thomas~P Fuglseth, Petter T{\o}ndel, and Thor~I Fossen.
\newblock Optimal constrained control allocation in marine surface vessels with
  rudders.
\newblock {\em Control Engineering Practice}, 16(4):457--464, 2008.

\bibitem{KirSteHor18}
Martin Kirchengast, Martin Steinberger, and Martin Horn.
\newblock Input matrix factorizations for constrained control allocation.
\newblock {\em IEEE Transactions on Automatic Control}, 63(4):1163--1170, 2018.

\bibitem{LavGib11}
Eugene Lavretsky and Travis~E Gibson.
\newblock Projection operator in adaptive systems.
\newblock {\em arXiv preprint arXiv:1112.4232}, 2011.

\bibitem{Lia10}
F~Liao, K-Y Lum, JL~Wang, and M~Benosman.
\newblock Adaptive control allocation for non-linear systems with internal
  dynamics.
\newblock {\em IET control theory \& applications}, 4(6):909--922, 2010.

\bibitem{PetBod06}
John~AM Petersen and Marc Bodson.
\newblock Constrained quadratic programming techniques for control allocation.
\newblock {\em IEEE Transactions on Control Systems Technology}, 14(1):91--98,
  2006.

\bibitem{Pod01}
Tarun~Kanti Podder and Nilanjan Sarkar.
\newblock Fault-tolerant control of an autonomous underwater vehicle under
  thruster redundancy.
\newblock {\em Robotics and Autonomous Systems}, 34(1):39--52, 2001.

\bibitem{Rao17}
M~Ehsan Raoufat, Kevin Tomsovic, and Seddik~M Djouadi.
\newblock Dynamic control allocation for damping of inter-area oscillations.
\newblock {\em IEEE Transactions on Power Systems}, 32(6):4894--4903, 2017.

\bibitem{SadChaZhaThe12}
Iman Sadeghzadeh, Abbas Chamseddine, Youmin Zhang, and Didier Theilliol.
\newblock Control allocation and re-allocation for a modified quadrotor
  helicopter against actuator faults.
\newblock {\em IFAC Proceedings Volumes}, 45(20):247--252, 2012.

\bibitem{She15}
Qiang Shen, Danwei Wang, Senqiang Zhu, and Eng~Kee Poh.
\newblock Inertia-free fault-tolerant spacecraft attitude tracking using
  control allocation.
\newblock {\em Automatica}, 62:114--121, 2015.

\bibitem{She17}
Qiang Shen, Danwei Wang, Senqiang Zhu, and Eng~Kee Poh.
\newblock Robust control allocation for spacecraft attitude tracking under
  actuator faults.
\newblock {\em IEEE Transactions on Control Systems Technology},
  25(3):1068--1075, 2017.

\bibitem{Sor11}
Asgeir~J S{\o}rensen.
\newblock A survey of dynamic positioning control systems.
\newblock {\em Annual reviews in control}, 35(1):123--136, 2011.

\bibitem{Sor17}
Mikkel Eske~N{\o}rgaard S{\o}rensen, S{\o}ren Hansen, Morten Breivik, and
  Mogens Blanke.
\newblock Performance comparison of controllers with fault-dependent control
  allocation for uavs.
\newblock {\em Journal of Intelligent \& Robotic Systems}, 87(1):187--207,
  2017.

\bibitem{Ste17}
Johannes Stephan and Walter Fichter.
\newblock Fast exact redistributed pseudoinverse method for linear actuation
  systems.
\newblock {\em IEEE Transactions on Control Systems Technology}, (99):1--8,
  2017.

\bibitem{SteLew15}
Brian~L Stevens, Frank~L Lewis, and Eric~N Johnson.
\newblock {\em Aircraft control and simulation: dynamics, controls design, and
  autonomous systems}.
\newblock John Wiley \& Sons, 2015.

\bibitem{Tag11}
Hamid~D Taghirad and Yousef~B Bedoustani.
\newblock An analytic-iterative redundancy resolution scheme for cable-driven
  redundant parallel manipulators.
\newblock {\em IEEE Transactions on Robotics}, 27(6):1137--1143, 2011.

\bibitem{TjoJoh08}
Johannes Tj{\o}nn{\aa}s and Tor~A Johansen.
\newblock Adaptive control allocation.
\newblock {\em Automatica}, 44(11):2754--2765, 2008.

\bibitem{TjoJoh10}
Johannes Tj{\o}nn{\aa}s and Tor~A Johansen.
\newblock Stabilization of automotive vehicles using active steering and
  adaptive brake control allocation.
\newblock {\em IEEE Transactions on Control Systems Technology},
  18(3):545--558, 2010.

\bibitem{Toh16}
Seyed~Shahabaldin Tohidi, Ali Khaki~Sedigh, and David Buzorgnia.
\newblock Fault tolerant control design using adaptive control allocation based
  on the pseudo inverse along the null space.
\newblock {\em International Journal of Robust and Nonlinear Control},
  26(16):3541--3557, 2016.

\bibitem{TohYil16}
Seyed~Shahabaldin Tohidi, Yildiray Yildiz, and Ilya Kolmanovsky.
\newblock Fault tolerant control for over-actuated systems: An adaptive
  correction approach.
\newblock In {\em American Control Conference (ACC), 2016}, pages 2530--2535.
  IEEE, 2016.

\bibitem{TohYil17}
Seyed~Shahabaldin Tohidi, Yildiray Yildiz, and Ilya Kolmanovsky.
\newblock Adaptive control allocation for over-actuated systems with actuator
  saturation.
\newblock volume~50, pages 5492--5497. Elsevier, 2017.

\bibitem{TohYil18}
Seyed~Shahabaldin Tohidi, Yildiray Yildiz, and Ilya Kolmanovsky.
\newblock Pilot induced oscillation mitigation for unmanned aircraft systems:
  An adaptive control allocation approach.
\newblock In {\em 2018 IEEE Conference on Control Technology and Applications
  (CCTA)}, pages 343--348. IEEE, 2018.

\bibitem{Vir94}
John Virnig and David Bodden.
\newblock Multivariable control allocation and control law conditioning when
  control effectors limit.
\newblock In {\em Guidance, Navigation, and Control Conference}, page 3609,
  1994.

\bibitem{Yil11b}
Yildiray Yildiz and Ilya Kolmanovsky.
\newblock Implementation of capio for composite adaptive control of
  cross-coupled unstable aircraft.
\newblock In {\em Infotech@ Aerospace 2011}, page 1460. 2011.

\bibitem{Yil11a}
Yildiray Yildiz and Ilya Kolmanovsky.
\newblock Stability properties and cross-coupling performance of the control
  allocation scheme capio.
\newblock {\em Journal of Guidance, Control, and Dynamics}, 34(4):1190--1196,
  2011.

\bibitem{YilKol10}
Yildiray Yildiz and Ilya~V Kolmanovsky.
\newblock A control allocation technique to recover from pilot-induced
  oscillations (capio) due to actuator rate limiting.
\newblock In {\em Proceedings of the 2010 American Control Conference}, pages
  516--523. IEEE, 2010.

\bibitem{YilKolAco11}
Yildiray Yildiz, Ilya~V Kolmanovsky, and Diana Acosta.
\newblock A control allocation system for automatic detection and compensation
  of phase shift due to actuator rate limiting.
\newblock In {\em Proceedings of the 2011 American Control Conference}, pages
  444--449. IEEE, 2011.

\bibitem{Zac09}
Luca Zaccarian.
\newblock Dynamic allocation for input redundant control systems.
\newblock {\em Automatica}, 45(6):1431--1438, 2009.

\bibitem{Zha08}
Youmin Zhang and Jin Jiang.
\newblock Bibliographical review on reconfigurable fault-tolerant control
  systems.
\newblock {\em Annual reviews in control}, 32(2):229--252, 2008.

\end{thebibliography}

\end{document}